\documentclass[11pt]{article}
\usepackage{longtable}
\usepackage{caption}
\usepackage{multirow,multicol}
\usepackage{epsf, graphicx}
\usepackage{latexsym,amsfonts,amsbsy,amssymb}
\usepackage{amsmath,amsthm}
\usepackage{float}
\usepackage{bm}
\usepackage[colorlinks,linkcolor=blue,anchorcolor=blue,citecolor=blue]{hyperref}

\usepackage[linesnumbered,lined,boxed,commentsnumbered]{algorithm2e}

\usepackage{cleveref}
\usepackage{indentfirst}
\usepackage{bm}
\usepackage{enumitem}

\allowdisplaybreaks[4]
\makeatletter

\DeclareMathOperator{\GRS}{GRS}
\DeclareMathOperator{\TGRS}{TGRS}
\DeclareMathOperator{\NGRS}{NGRS}
\DeclareMathOperator{\EGRS}{EGRS}
\DeclareMathOperator{\MGRS}{MGRS}
\DeclareMathOperator{\False}{False}
\DeclareMathOperator{\True}{True}

\DeclareMathOperator{\EMGRS}{EMGRS}
\DeclareMathOperator{\Diag}{Diag}
\@addtoreset{equation}{section} \makeatother
\newtheorem{Theorem}{Theorem}[section]
\newtheorem{Lemma}{Lemma}[section]
\newtheorem{Corollary}{Corollary}[section]
\newtheorem{Remark}{Remark}[section]
\newtheorem{Definition}{Definition}[section]
\newtheorem{Proposition}{Proposition}[section]
\newtheorem{Problem}{Problem}[section]
\newtheorem{Example}{Example}[section]
\makeatletter \@addtoreset{equation}{section} \makeatother
\begin{document}
	
	\title{
        New Constructions of Non-GRS MDS Codes,
        Recovery and Determination Algorithms for GRS Codes
    \let\thefootnote\relax\footnote{E-mail addresses: wanggdmath@163.com (G. Wang), hwliu@ccnu.edu.cn (H. Liu), luojinquan@ccnu.edu.cn (J. Luo)}
    }
	\author{Guodong Wang,
    Hongwei Liu,
    Jinquan Luo
    }
	\date{\small
    School of Mathematics and Statistics,
		Central China Normal University,
		Wuhan,  430079, China
	}
	\maketitle
	{\noindent\small{\bf Abstract:}
    In this paper, we propose a new method for constructing a class of non-GRS MDS codes.
The lengths of these codes can reach up to $\frac{q+3}{2}$ (for finite fields of odd characteristic) and $\frac{q+4}{2}$ (for even characteristic), respectively.
Owing to their special structure,
we can use the Cauchy matrix method to obtain the necessary and sufficient conditions for these codes to be MDS codes and non-GRS MDS codes.
Additionally, the inequivalence between these codes and twisted GRS codes is analyzed.
Furthermore, we analyze the relationships among several existing classes of codes used for constructing non-GRS MDS codes,
propose explicit constructions,
and discuss the lengths of non-GRS MDS codes based on these constructions.
Finally, we design two efficient algorithms to address two main problems in GRS code research,
i.e., determining whether an unknown code $C$ is a GRS code from its generator matrix $G$,
and recovering the key vectors $\bm{\alpha}$ and $\bm{v}$ such that $C = \GRS_{n,k}(\bm{\alpha}, \bm{v})$ if $C$ is indeed a GRS code.
A computational complexity comparison of the proposed algorithms ($O(nk+n)$) with that of the Sidelnikov-Shestakov attack (exceeding $O(qk^2n+qk^3)$)
shows that our methods offer superior computational efficiency.
}
	\vspace{1ex}

	
	
	\section{Introduction}
    A linear code with parameters $[n, k, d]$ is referred to as a {\em maximum distance separable} (MDS) code if it meets the Singleton bound,
    i.e., $d = n - k + 1$. When $d = n - k$, such codes are called {\em almost-MDS} codes.
    Due to their excellent distance properties, MDS codes have been widely studied.
    Closely related to MDS codes is the famous MDS conjecture,
    which states that for an $[n, k, d]$ MDS code with $d \ge 3$, the length $n$ must satisfy $n \le q + 1$,
    except in the special case where $q = 2^h$ and $k \in \{3, q - 1\}$, in which case $n \le q + 2$.
    For prime $q$, the conjecture is proven by Ball \cite{Ball1}, in particular, if $4 \le k \le q - 3$, then any $k$-dimensional linear MDS code must satisfy $n \le q + 1$.
    Originally proposed by Segre \cite{Segre} in the context of finite projective geometry (see also \cite{Ball2}),
    In particular, {\em generalized Reed-Solomon} (GRS) codes are a class of MDS codes.
    Based on GRS codes, various types of MDS codes have been constructed, such as {\em self-orthogonal} \cite{Ball3, Chen2},
    {\em self-dual} MDS codes \cite{Fang, Lebed, Niu, Sok}, and  {\em linear complementary dual} (LCD) MDS codes \cite{Carlet}.

    When an MDS code is not equivalent to any GRS code, it is referred to as a {\em non-GRS MDS} codes.
    In 1989, Roth and Lempel \cite{Roth} constructed non-GRS MDS codes by extending extended GRS codes.
    Later, Beelen et al. \cite{Beelen1} introduced a new class of codes,
    called {\em twisted Reed-Solomon} (TRS) codes,
    presenting general constructions of non-GRS MDS codes.
    Following this line of research, subsequent works have investigated their dual codes, self-dual codes \cite{Ding, Guo, Huang1, Sui2, Sui4}, LCD codes \cite{Huang2},
    hulls (the intersections of these codes and their duals) \cite{Wu3}, decoding algorithms \cite{Sui3, Sun1, Wang},
    and multiple-twists GRS codes \cite{Ding, Hu, Meena, Zhao}.
    Additionally, Chen \cite{Chen} constructed numerous non-GRS MDS codes based on algebraic curves.
    Recently, Li et al. constructed a class of cyclic MDS codes that are not equivalent to Reed-Solomon codes \cite{Li}.
    Wu et al. \cite{Wu4} conducted research on whether the extension of an MDS code is also an MDS code.
    Liu et al. adopted a new method to construct non-GRS MDS codes, referred to as {\em column-twisted Reed-Solomon} codes \cite{Liu},
    which over a finite field $\mathbb{F}_{q}$ (with $q$ an odd prime power) can achieve code lengths up to $\frac{q+3}{2}$.
    Han and Ren \cite{Han2} further used elliptic curves to show that MDS codes can attain length $\frac{q+1}{2} + \lfloor\sqrt{q}\rfloor$ under certain conditions on code dimension $k$.

    The {\em Schur product} method \cite{Mirandola} is commonly used to determine whether a constructed MDS code is non-GRS
    and has been widely adopted in \cite{Beelen1, Chen, Hu, Jin, Li, Liyang, Liyang2, Liu, Sui1, Zhu}.
    However, the Schur product method has a limitation: it requires $k < \frac{n+1}{2}$,
    where $k$ and $n$ denote the code dimension and length, respectively.
    Earlier, Roth and Lempel (1989) proposed the {\em Cauchy matrix} method ([Lemma 7]\cite{Roth2}, \cite{Roth3}) to characterize the necessary and sufficient conditions for a code to be a GRS code.
    Unlike the Schur product method, the Cauchy matrix method is free from this constraint. 
    A typical class of non-GRS MDS codes identified by this method is the {\em Roth-Lempel} codes \cite{Roth}.
    It is obvious that the codes obtained by extending the columns of Roth-Lempel codes are also non-GRS codes \cite{Liang,Wu5}.
    
    Based on the construction idea of TRS, Jin et al. \cite{Jin} constructed non-GRS MDS codes by using subcodes of GRS codes,
    with the specific method of removing one row from the standard GRS generator matrix.
    The Schur product of such codes is more straightforward to compute.
    We construct another code that shares the same MDS condition as that in \cite{Jin}, but has fewer restrictions on the code dimension.
    Moreover, this code is not equivalent to the construction proposed in \cite[Section VI]{Jin}.
    In subsequent work, Li et al. \cite{Liyang} consider a class of codes that are subcodes of extended GRS codes,
    whose generator matrices are obtained by deleting one specific column from the extended GRS codes.
    Abdukhalikov et al. \cite{Abdukhalikov} adopted a similar approach.
    They investigated subcodes of extended GRS codes and subcodes of Roth-Lempel codes to construct non-GRS MDS codes.

    In this paper, we make extensive use of the Cauchy matrix method.
    We define a class of codes called {\em modified GRS} codes,
    with a generator matrix that differs from that of a GRS code in only one entry.
    Owing to this special structure, the Cauchy matrix method enables us to derive necessary and sufficient conditions for both these codes and their extended versions to be non-GRS MDS codes,
    without restrictions on the code dimension.
    Furthermore, we have shown that the (extended) modified GRS codes are not equivalent to the twisted GRS codes $C^{n,k}_{\bm{\alpha, 1,s,\eta}}$ \cite[Definition 1]{Beelen1} where $0 \le s \le k-1$.
    Additionally, decoding these modified GRS codes requires only minor modifications to existing algorithms \cite{Sui3, Sun1, Wang}.
    We also provide a concise proof showing that Roth-Lempel codes are non-GRS codes.

Currently, research on the interrelationships among (extended) modified GRS codes, (extended) column-twisted GRS codes \cite{Liu},
Roth-Lempel codes \cite{Roth}, and twisted GRS codes \cite{Beelen1} remains relatively limited.
To address this gap, this paper analyzes the interrelationships among these code classes.
We also propose numerous specific constructions for non-GRS MDS codes,
and discuss the lengths of non-GRS MDS codes based on these constructions.

On the other hand, given a generator matrix $G$ of an unknown code $C$,
how can we efficiently determine whether $C$ is a GRS code?
This is a main problem frequently encountered in constructing non-GRS MDS codes.
Furthermore, if $C$ is indeed a GRS code,
how can we recover its key information using the generator matrix $G$, specifically,
finding vectors $\bm{\alpha}$ and $\bm{v}$ such that $C = \GRS_{n,k}(\bm{\alpha}, \bm{v})$?
Notably, the problems of GRS code identification and information recovery are also relevant to the field of cryptography.
For example, the Niederreiter scheme \cite{Niederreiter} is a public-key cryptosystem based on GRS codes,
which was successfully attacked by Sidelnikov and Shestakov \cite{Sidelnikov} through what is now known as the 
{\em Sidelnikov-Shestakov attack} method.
To address the two aforementioned problems, i.e., GRS code identification and GRS code information recovery,
this paper presents two efficient algorithms.
A complexity comparison with the Sidelnikov-Shestakov attack demonstrates that our algorithms achieve higher computational efficiency.

It is known that if the punctured code and shortened code of a code at a specific position are both MDS codes,
then the original code must also be MDS.
This observation motivates the following question:
If the punctured code and shortened code of a code at a given position are both GRS codes,
does the original code necessarily have to be a GRS code?
We present a counterexample showing that the answer is negative.

Paper organization:
Section 2 introduces the necessary preliminaries.
Section 3 proposes a new class of codes for constructing non-GRS MDS codes, and derives necessary and sufficient conditions for this code class to be MDS codes or non-GRS MDS codes.
Section 4 analyzes the interrelationships between several classes of codes used to construct non-GRS MDS codes.
Section 5 provides explicit constructions and their maximum length to be non-GRS MDS codes.
Section 6 describes the proposed algorithms for identifying and recovering GRS codes,
analyzes their complexity, and compares their performance with the Sidelnikov-Shestakov attack.
\footnote{The Magma code can be found at \url{https://github.com/1wangguodong/Recovery-and-Determination-Algorithms-for-GRS-Codes}}

\section{Preliminaries}
Let $\mathbb{F}_q$ be the finite field of order $q$, where $q$ is a power of a prime $p$.
Unless otherwise specified, we always assume $\alpha_1, \ldots, \alpha_n$ are distinct elements of $\mathbb{F}_q$ and $v_1, \ldots, v_n, v_{n+1}$ are nonzero elements of $\mathbb{F}_q$.
We denote $\bm{\alpha} = (\alpha_1, \ldots, \alpha_t)$ and $\bm{v} = (v_1, \ldots, v_s)$,
where $t$ and $s$ are determined by specific scenarios: $t$ is either $n-1$ or $n$, and $s$ is either $n$ or $n+1$.

Let $C$ be an $[n,k]$ code and let $\mathcal{N}$ be a subset of $\{1,2,\ldots,n\}$.
The {\em punctured code} of $C$ on $\mathcal{N}$ is the code obtained by deleting all the coordinates in $\mathcal{N}$ for each codeword of $C$.
Let
\[
    C(\mathcal{N})=\{ \bm{c}=(c_1,\ldots,c_i,\ldots,c_n) \in C \,|\,  c_i=0 \ \text{for all}\ i\in \mathcal{N}\}
\]
be the subcode of $C$.
The punctured code of $C(\mathcal{N})$ on $\mathcal{N}$ is called a {\em shortened code} of $C$.
The {\em dual code} $C^{\bot}$ of $C$ is defined as $C^{\bot} = \{\bm{x} \in \mathbb{F}_q^n \,|\, \langle \bm{x},\bm{y} \rangle =0$ for all $\bm{y} \in C\}$,
where $\langle \bm{x},\bm{y} \rangle=\sum_{i=1}^nx_iy_i$ denotes the {\em Euclidean (standard) inner product}.

\begin{Definition}
For $0 \le k \le n$, the {\em generalized Reed-Solomon} (GRS) code is as follows:
$$
    \GRS_{n,k}(\bm{\alpha}, \bm{v})=\left\{\left(v_1 f\left(\alpha_1\right), v_2 f\left(\alpha_2\right), \ldots, v_n f\left(\alpha_n\right)\right) \mid f(x) \in \mathbb{F}_q[x]_{k}\right\},
$$
where $\mathbb{F}_q[x]_{k}$ denotes the set of polynomials in $\mathbb{F}_q[x]$ of degree less than $k$,
which is a vector space of dimension $k$ over $\mathbb{F}_q$.
\end{Definition}

A GRS code $\GRS_{n,k}(\bm{\alpha}, \bm{v})$
is an $[n, k, n-k+1]$ linear code over $\mathbb{F}_q$, which has a generator matrix
$$
    \begin{pmatrix}
    1 & \cdots & 1 \\
    \alpha_1 & \cdots & \alpha_n \\
    \vdots & & \vdots \\
    \alpha_1^{k-1} & \cdots & \alpha_n^{k-1}
    \end{pmatrix}\Diag(v_1,v_2,...,v_n),
$$
where $\Diag(v_1,v_2,...,v_n)$ denotes the diagonal matrix.
\begin{Remark}\label{GRSword}
It is easy to see that $\GRS_{n,k}(\bm{\alpha}, \bm{v})$ consists exactly of those codewords $\bm{c}\in\mathbb{F}_q^n$ for which a (unique) polynomial $f_{\bm{c}}\in\mathbb{F}_q[x]$ of degree at most $k - 1$ exists such that
\[
    \bm{c}=(v_1f_{\bm{c}}(\alpha_1),v_2f_{\bm c}(\alpha_2),\ldots,v_nf_{\bm c}(\alpha_n)).
\]
We call $f_{\bm c}$ the polynomial associated with $\bm{c}$.
\end{Remark}

It is well-known that the dual of a GRS code is also a GRS code:
\begin{Proposition}\label{GRSDual}
Let $\bm{\alpha},\bm{v}$ be defined as above and $\bm{u} =(u_1,\ldots u_n)$, where $u_i =v_i^{-1}\prod_{j\neq i}(\alpha_i - \alpha_j)^{-1}$.
Then, the dual code of $\GRS_{n,k}(\bm{\alpha}, \bm{v})$ is given by
\[
    \GRS_{n,k}(\bm{\alpha},\bm{v})^{\perp}=\GRS_{n,n - k}(\bm{\alpha}, \bm{u}).
\]
\end{Proposition}

\begin{Definition}
    For $0 \le k \le n$,
    assume $\alpha_1, \ldots, \alpha_n$ are distinct elements of  $\mathbb{F}_q\cup \{\infty\}$, and $v_1, \ldots, v_n$ are nonzero elements of $\mathbb{F}_q$,
    denoted by $\bm{\alpha}=(\alpha_1, \ldots, \alpha_n)$ and $\bm{v}=(v_1, \ldots, v_n)$.
    The {\em extended generalized Reed-Solomon code} (extended GRS) is defined as:
$$
    \EGRS_{n,k}(\bm{\alpha}, \bm{v})=\left\{\left(v_1 f\left(\alpha_1\right), v_2 f\left(\alpha_2\right), \ldots, v_n f\left(\alpha_n\right)\right) \mid f(x) \in \mathbb{F}_q[x]_{k}\right\},
$$
    where $f(\infty)=f_{k-1}$, and $f_{k-1}$ is the coefficient of $x^{k-1}$ in $f(x)$.
\end{Definition}

An extended GRS code $\EGRS_{n,k}(\bm{\alpha}, \bm{v})$ with $\alpha_i=\infty$
is an $[n, k, n-k+1]$ linear code over $\mathbb{F}_q$, which has a generator matrix
$$
    G=\begin{pmatrix}
    1 & \cdots & 1 & 0 & 1 & \cdots & 1 \\
    \alpha_1 & \cdots & \alpha_{i-1} & 0 & \alpha_{i+1} & \cdots &  \alpha_n \\
    \vdots & & \vdots &  & \vdots & \cdots & \vdots \\
    \alpha_1^{k-1} & \cdots & \alpha_{i-1}^{k-1} & 1 & \alpha_{i+1}^{k-1} & \cdots & \alpha_n^{k-1}
    \end{pmatrix}\Diag(v_1,v_2,...,v_n).
$$

According to the definition, when $\alpha_j\neq\infty$ for all $j = 1,\ldots,n$, an extended GRS code is a GRS code.
We define $\frac{1}{\infty} = 0$ and $\frac{1}{0} = \infty$.
A useful property about extended GRS codes is given below.

\begin{Proposition}\label{trans}
Let $a,b,c \in \mathbb{F}_q$ with $a,c \neq 0$. Then:
\begin{enumerate}
    \item $\EGRS_{n,k}((\alpha_1,...,\alpha_n), (v_1,...,v_n)) = \EGRS_{n,k}((a\alpha_1+b,...,a\alpha_n+b), (cv_1,...,cv_n))$.
    \item $\EGRS_{n,k}((\alpha_1,...,\alpha_n), (v_1,...,v_n)) = \EGRS_{n,k}((\frac{1}{\alpha_1},...,\frac{1}{\alpha_n}), (v_1',...,v_n'))$,
    where $$v_j'=\begin{cases}
        v_j, & \text{if} \ \alpha_j=0 \ \text{or} \ \alpha_j=\infty,  \\
        v_j\alpha_j^{k-1}, &\text{otherwise},
    \end{cases}$$
    for $j=1,...,n$.
\end{enumerate}
\end{Proposition}

\begin{Remark}\label{remarkeq}
    The above transformation implies that GRS codes and extended GRS codes are equivalent when the code length is less than $q+1$.
\end{Remark}

For an MDS code, there are two commonly used methods to determine whether it is equivalent to a GRS code:
\begin{enumerate}
    \item the {\em Cauchy matrix} method,
    \item the {\em Schur product} method.
\end{enumerate}

To the best of our knowledge, for most non-GRS MDS codes constructed so far,
the Schur product method has been widely used to verify their non-equivalence to GRS codes \cite{Beelen1, Chen, Hu, Jin, Li, Liyang, Liyang2, Liu, Sui1, Zhu}.
However, this method has an important limitation: it requires that $ k < \frac{n+1}{2} $,
where $k$ denotes the code dimension and $n$ the code length.
In contrast, the Cauchy matrix method is free of such a restriction.

In this paper, we employ the Cauchy matrix method to prove that the proposed MDS codes are not equivalent to any GRS code.
In what follows, we introduce only the Cauchy matrix method;
for further details on the Schur product method, we refer the reader to \cite{Mirandola}.

\begin{Lemma}\cite{Roth2,Roth3}\label{cauchy1}
    A matrix of the form $G = [I \mid A]$ generates an $[n,k]$ GRS code if and only if $A=[a_{i,j}]$ is a Cauchy matrix, i.e.,
\[
    a_{i,j}=\frac{c_id_j}{x_i+y_j}, 0 \le i \le k-1, 0 \le j \le n-k-1,
\]
where the $x_i$ are distinct elements of $\mathbb{F}_q$, the $y_j$ are distinct elements of $\mathbb{F}_q$,
$x_i+y_j \neq 0$ for all $i$ and $j$, and $c_i,d_j \neq 0$.
\end{Lemma}

\begin{Lemma}\cite{Roth2}\label{cauchy2}
    Let $ \mathcal{C} $ be a linear code with generator matrix of the form $G = [I \mid A]$.
Then $A=[a_{i,j}]$ is a Cauchy matrix if and only if, for $A' \in \mathbb{F}_q^{k \times (n - k)} $ with $A' = [a_{i,j}^{-1}]$,
\begin{enumerate}
    \item all entries of $A$ are nonzero,
    \item all $ 2 \times 2 $ minors of $A'$ are nonzero, and
    \item all $ 3 \times 3 $ minors of $A'$ are zero.
\end{enumerate}
\end{Lemma}

\begin{Remark}
Lemmas \ref{cauchy1} and \ref{cauchy2} also hold in two cases:
(1) when the identity matrix $I$ in the generator matrix $G = [I \mid A]$ is replaced by an invertible diagonal matrix; and
(2) when the matrix $G$ is right-multiplied by an invertible diagonal matrix.
\end{Remark}

\section{New class of non-GRS MDS codes}

In this section, we propose two classes of codes, namely,
the modified GRS codes and the extended modified GRS codes.
We provide the necessary and sufficient conditions for these codes to be MDS and non-GRS. 

\subsection{Modified GRS codes}

\begin{Definition}\label{def1}
    For $1 \le k \le n$ and $1 \le t \le k-1$, we define the {\em modified generalized Reed-Solomon} (modified GRS) code as
    $$\MGRS_{n,k}(\bm{\alpha}, \bm{v}, \eta, t)=\left\{\left(v_1 f\left(\alpha_1\right), \ldots,v_{n-1}f\left(\alpha_{n-1}\right), v_n \left(f\left(0\right)+\eta f_t\right)\right) \mid f(x) \in \mathbb{F}_q[x]_{k}\right\},$$
    where $\eta \in \mathbb{F}_q$ and $f_t$ denotes the coefficient of $x^t$ in $f(x)$.
    Note that if $\eta=0$ and $\alpha_i \neq 0$ for $i=1,...,n-1$, then $\MGRS_{n,k}(\bm{\alpha}, \bm{v}, \eta, t)$ is the GRS code.
\end{Definition}

Clearly, $\MGRS_{n,k}(\bm{\alpha}, \bm{v}, \eta, t)$ has the following generator matrix:
\[
    G = \begin{bmatrix} 
    1 & 1 & \cdots & 1 & 1 \\
    \alpha_1 & \alpha_2 & \cdots & \alpha_{n-1} & 0 \\
    \vdots & \vdots & \cdots & \vdots & \vdots \\
    \alpha_1^{t - 1} & \alpha_2^{t - 1} & \cdots & \alpha_{n-1}^{t - 1} & 0 \\
    \alpha_1^{t} & \alpha_2^{t} & \cdots & \alpha_{n-1}^{t} & \eta \\
    \alpha_1^{t + 1} & \alpha_2^{t + 1} & \cdots & \alpha_{n-1}^{t + 1} & 0 \\
    \vdots & \vdots & \cdots & \vdots & \vdots \\
    \alpha_1^{k - 1} & \alpha_2^{k - 1} & \cdots & \alpha_{n-1}^{k - 1} & 0
    \end{bmatrix}\Diag(v_1,v_2,...,v_{n}).
\]

\begin{Remark}
    The code $\MGRS_{n,k}(\bm{\alpha}, \bm{v}, \eta, t)$ is either an MDS or an almost-MDS code.
\end{Remark}

\begin{Theorem}\label{LMDS}
    Assume the notations are given as above.
Then the code $\MGRS_{n,k}(\bm{\alpha}, \bm{v}, \eta, t)$ is MDS if and only if $\eta\pi_t \neq  (-1)^{k}\prod_{s=1}^{k - 1} \alpha_{i_s}$,
where $\{i_1, i_2, \dots, i_{k-1}\}$ is an arbitrary $(k-1)$-element subset of $\{1, 2, \dots, n-1\}$ and
$$\pi_s = (-1)^{k-1-s} \cdot \sum_{1 \leq j_1 < j_2 < \cdots < j_{k-1-s} \leq k-1} \alpha_{i_{j_1}}\alpha_{i_{j_2}}\cdots \alpha_{i_{j_{k-1-s}}} \ \text{for} \ s=0,...,k-1.$$
\begin{proof}
    For any $(k-1)$-element subset $\{i_1, i_2, \dots, i_{k-1}\}$ of $\{1, 2, \dots, n-1\}$, we have
    $$
        \prod_{j=1}^{k-1}(x-a_{i_j})=\pi_{k-1}x^{k-1}+\pi_{k-2}x^{k-2}+\cdots + \pi_1x+\pi_0,
    $$
where $\pi_s = (-1)^{k-1-s} \cdot \sum_{1 \leq j_1 < j_2 < \cdots < j_{k-1-s} \leq k-1} \alpha_{s_{j_1}}\alpha_{i_{j_2}}\cdots \alpha_{i_{j_{k-1-s}}}$
for $s=0,...,k-1$.
We only need to ensure that the following determinant is nonzero.
\begin{align*}
    &\det\left(\begin{bmatrix}
1 & 1 & \cdots & 1 & 1 \\
\alpha_{i_1} & \alpha_{i_2} & \cdots & \alpha_{i_{k-1}} & 0 \\
\vdots & \vdots & \cdots & \vdots & \vdots \\
\alpha_{i_1}^t & \alpha_{i_2}^t & \cdots & \alpha_{i_{k-1}}^t & \eta \\
\vdots & \vdots & \cdots & \vdots & \vdots \\
\alpha_{i_1}^{k - 1} & \alpha_{i_2}^{k - 1} & \cdots & \alpha_{i_{k-1}}^{k - 1} & 0
\end{bmatrix}\right) \\
=&\det\left(\begin{bmatrix}
1 & 1 & \cdots & 1 & 1 \\
\alpha_{i_1} & \alpha_{i_2} & \cdots & \alpha_{i_{k-1}} & 0 \\
\vdots & \vdots & \cdots & \vdots & \vdots \\
\alpha_{i_1}^{k - 2} & \alpha_{i_2}^{k - 2} & \cdots & \alpha_{i_{k-1}}^{k - 2} & 0 \\
\alpha_{i_1}^{k - 1} & \alpha_{i_2}^{k - 1} & \cdots & \alpha_{i_{k-1}}^{k - 1} & 0
\end{bmatrix}\right)+\det\left(\begin{bmatrix}
1 & 1 & \cdots & 1 & 0 \\
\alpha_{i_1} & \alpha_{i_2} & \cdots & \alpha_{i_{k-1}} & 0 \\
\vdots & \vdots & \cdots & \vdots & \vdots \\
\alpha_{i_1}^t & \alpha_{i_2}^t & \cdots & \alpha_{i_{k-1}}^t & \eta \\
\vdots & \vdots & \cdots & \vdots & \vdots \\
\alpha_{i_1}^{k - 1} & \alpha_{i_2}^{k - 1} & \cdots & \alpha_{i_{k-1}}^{k - 1} & 0
\end{bmatrix}\right)\\
=&\prod_{\substack{1 \leq u < v \leq k - 1}} (\alpha_{i_v} - \alpha_{i_u}) \cdot \left( \prod_{s=1}^{k - 1} (-\alpha_{i_s}) + \eta\pi_t \right).
\end{align*}
Then the  code $\MGRS_{n,k}(\bm{\alpha}, \bm{v}, \eta, t)$ is MDS if and only if $\eta\pi_t \neq (-1)^k\prod_{s=1}^{k - 1} \alpha_{i_s}$.
\end{proof}
\end{Theorem}

\begin{Corollary}\label{MDS1}
    The code $\MGRS_{n,k}(\bm{\alpha}, \bm{v}, \eta, k-1)$ is MDS if and only if $\eta \neq  (-1)^k\prod_{s=1}^{k - 1} \alpha_{i_s}$,
where $\{i_1, i_2, \dots, i_{k-1}\}$ is an arbitrary $(k-1)$-element subset of $\{1, 2, \dots, n-1\}$.
\begin{proof}
   By Theorem \ref{LMDS}, for any $(k-1)$-element subset $\{i_1, i_2, \dots, i_{k-1}\}$ of $\{1, 2, \dots, n-1\}$, 
the code $\MGRS_{n,k}(\bm{\alpha}, \bm{v}, \eta, k-1)$ is MDS if and only if $\eta \neq (-1)^k\prod_{s=1}^{k - 1} \alpha_{i_s}$, since $\pi_{k-1} = 1$.
\end{proof}
\end{Corollary}

For any  element $a \in \mathbb{F}_q$, we define $\infty + a = \infty$.
Then we have:
\begin{Corollary}\label{MDS2}
    The code $\MGRS_{n,k}(\bm{\alpha}, \bm{v}, \eta, 1)$ is MDS if and only if $\frac{1}{\eta} \neq \sum_{s=1}^{k-1}\frac{1}{\alpha_{i_s}}$,
where $\{i_1, i_2, \dots, i_{k-1}\}$ is an arbitrary $(k-1)$-element subset of $\{1, 2, \dots, n-1\}$.
\begin{proof}
        For any $(k-1)$-element subset $\{i_1, i_2, \dots, i_{k-1}\}$ of $\{1, 2, \dots, n-1\}$, 
    by Theorem \ref{LMDS}, the code $\MGRS_{n,k}(\bm{\alpha}, \bm{v}, \eta, 1)$ is MDS if and only if
    $\eta\pi_1 \neq (-1)^k\prod_{s=1}^{k - 1} \alpha_{i_s}$
    (i.e., $\frac{1}{\eta} \neq \sum_{s=1}^{k-1}\frac{1}{\alpha_{i_s}}$).
\end{proof}
\end{Corollary}

The literatures \cite{Abdukhalikov, Jin} discusses a class of codes as follows.
Let $C^{n,k}_{\bm{\alpha},t}$ be a linear code over $\mathbb{F}_q$ with the generator matrix given by
\[
    \begin{bmatrix} 
    1 & 1 & \cdots & 1 \\
    \alpha_1 & \alpha_2 & \cdots & \alpha_{n} \\
    \vdots & \vdots & \cdots & \vdots  \\
    \alpha_1^{t - 1} & \alpha_2^{t - 1} & \cdots & \alpha_{n}^{t-1}\\
    \alpha_1^{t + 1} & \alpha_2^{t + 1} & \cdots & \alpha_{n}^{t+1}\\
    \vdots & \vdots & \cdots & \vdots  \\
    \alpha_1^{k} & \alpha_2^{k} & \cdots & \alpha_{n}^{k}
    \end{bmatrix},
\]
where $1\le t < k-1$. It is obvious that the code $C^{n,k}_{\bm{\alpha},t}$ is a $k$-dimensional subcode of $\GRS_{n,k+1}(\bm{\alpha}, \bm{1})$.
Moreover, its Schur product is easy to compute, and sufficient conditions for $C^{n,k}_{\bm{\alpha},t}$ to be a non-GRS MDS code are provided \cite[Proposition VI.1]{Jin}.
Next, we briefly discuss another class of codes related to this family codes.

Let $D^{n,k}_{\bm{\alpha},t}$ be a linear code over $\mathbb{F}_q$ with generator matrix as follows
\[
    \begin{bmatrix} 
    1 & 1 & \cdots & 1 & 0 \\
    \alpha_1 & \alpha_2 & \cdots & \alpha_{n-1} & 0 \\
    \vdots & \vdots & \cdots & \vdots & \vdots \\
    \alpha_1^{t - 1} & \alpha_2^{t - 1} & \cdots & \alpha_{n-1}^{t - 1} & 0 \\
    \alpha_1^{t} & \alpha_2^{t} & \cdots & \alpha_{n-1}^{t} & 1 \\
    \alpha_1^{t + 1} & \alpha_2^{t + 1} & \cdots & \alpha_{n-1}^{t + 1} & 0 \\
    \vdots & \vdots & \cdots & \vdots & \vdots \\
    \alpha_1^{k-1} & \alpha_2^{k-1} & \cdots & \alpha_{n-1}^{k-1} & 0
    \end{bmatrix},
\]
where $1\le t < k-1$.

It is easy to see that the shortened code of $D^{n,k}_{\bm{\alpha},t}$ at the $n$-th coordinate position is the code $C^{n-1,k-1}_{\bm{\alpha},t}$. 
Moreover, $D^{n,k}_{\bm{\alpha},t}$ and $C^{n-1,k-1}_{\bm{\alpha},t}$ satisfy the same MDS condition.
More precisely, an MDS $D^{n,k}_{\bm{\alpha},t}$ code yields an MDS $C^{n-1,k-1}_{\bm{\alpha},t}$ code, and vice versa. 
Below, we provide a characterization of the code $D^{n,k}_{\bm{\alpha},t}$ being a non-GRS code.

For convenience, let $P(x)= \prod_{j=1}^k(x-\alpha_j)$ and let $f_i(x)=\frac{P(x)}{x-\alpha_i}$, for $i= 1,...,k$.
We denote $\sigma_{h,s}$ as the coefficient of $x^s$ in $f_h(x)$, and $\sigma_{P,s}$ as the coefficient of $x^s$ in $P(x)$.
\begin{align*}
    f_h(x)=&x^{k-1}+\sigma_{h,k-2}x^{k-2}+\sigma_{h,k-3}x^{k-3}+\cdots+\sigma_{h,1}x+\sigma_{h,0} \\
    -\alpha_hf_h(x)=&-\alpha_hx^{k-1}-\alpha_h\sigma_{h,k-2}x^{k-2}-\alpha_h\sigma_{h,k-3}x^{k-3}+\cdots-\alpha_h\sigma_{h,1}x-\alpha_h\sigma_{h,0} \\
    P(x)=&x^k+(\sigma_{h,k-2}-\alpha_h)x^{k-1}+(\sigma_{h,k-3}-\alpha_h\sigma_{h,k-2})x^{k-2}+(\sigma_{h,k-4}-\alpha_h\sigma_{h,k-3})x^{k-3}\\
    &+\cdots+(\sigma_{h,1}-\alpha_h\sigma_{h,2})x^2+(\sigma_{h,0}-\alpha_h\sigma_{h,1})x-\alpha_h\sigma_{h,0}.
\end{align*}
It follows that there is a relationship between the coefficients of $P(x)$ and $f_h(x)$ as follows.
\begin{align*}
    \sigma_{h,0} &= \sigma_{P,0}/(-\alpha_h) \\ 
    \sigma_{h,1} &= (\sigma_{P,1}-\sigma_{h,0})/(-\alpha_h) \\ 
        &\vdots \\
    \sigma_{h,k-2} &= (\sigma_{P,k-2}-\sigma_{h,k-3})/(-\alpha_h) \\
    \sigma_{h,k-1} &=1.
\end{align*}
Thus
\begin{align}
    \sigma_{i,s}=-\frac{1}{\alpha_i}\sum_{j=0}^{s}\frac{\sigma_{P,j}}{\alpha_i^{s-j}} \ \text{for} \ i=1,...,k, s=0,...,k-2.
\end{align}

\begin{Theorem}\label{SimNGRS}
    Assume that $3 \le k \le n-3$ and $0 < t < k-1$. Then, the code $D^{n,k}_{\bm{\alpha},t}$ is a non-GRS code.
\begin{proof}
    Without loss of generality, assume that $\alpha_1,\dots,\alpha_k$ are all nonzero elements.
    The code $D^{n,k}_{\bm{\alpha},t}$ has a generator matrix $G = [\Diag(f_1(\alpha_1),f_2(\alpha_2),...,f_k(\alpha_k)) \mid A]$,
    \begin{align*}
        A = \begin{bmatrix}
            f_1(\alpha_{k+1}) & f_1(\alpha_{k+2}) & \cdots & f_1(\alpha_{n-1}) & \sigma_{1,t} \\
            f_2(\alpha_{k+1}) & f_2(\alpha_{k+2}) & \cdots & f_2(\alpha_{n-1}) & \sigma_{2,t}\\
            \vdots & \vdots & \cdots & \vdots & \vdots \\
             &  & \cdots &  &  \\
            f_{k}(\alpha_{k+1}) & f_{k}(\alpha_{k+2}) & \cdots & f_{k}(\alpha_{n-1}) & \sigma_{k,t}
            \end{bmatrix}.
    \end{align*}

    (1) Proof by contradiction.
    Assume $D^{n,k}_{\bm{\alpha},t}$ is a GRS code. Thus, all entries of $A=[a_{i,j}]$ are nonzero.
    By Lemmas \ref{cauchy1} and \ref{cauchy2}, any three columns of the matrix $A' = [a_{i,j}^{-1}]$ are linearly dependent.
    This implies that the vectors
    \begin{align*}
        (1/\sigma_{1,t},...,1/\sigma_{k,t}), (1,...,1),\ \text{and} \ (\alpha_1,...,\alpha_k)
    \end{align*}
    are linearly dependent.
    Thus, there exist $a,b \in \mathbb{F}_q$ (not both zero) such that $1/\sigma_{i,t} = a + b\alpha_i$ for all $1 \le i \le k$,
    which is equivalent to $a\sigma_{i,t} + b\alpha_i\sigma_{i,t} - 1 = 0$.

Define the polynomial 
$$
    m(x) = -ax\sum_{j=0}^{t}\sigma_{P,j}x^{t-j} - b\sum_{j=0}^{t}\sigma_{P,j}x^{t-j} - 1.
$$
When $0 < t < k-1$, $m(x)$ is a nonzero polynomial of degree $t+1$ (or $t$) by $\sigma_{P,0}\ne 0$.
Therefore, $m(x)$ has at most $t+1$ zeros.
However, $t+1 < k$, which contradicts the fact that $m(1/\alpha_i) = 0$ for all $i = 1,\dots,k$ (derived from $a\sigma_{i,t} + b\alpha_i\sigma_{i,t} - 1 = 0$).
This completes the proof.
\end{proof}
\end{Theorem}

\begin{Remark}
\begin{enumerate}
    \item For the case of non-GRS MDS codes, compared with the code $C^{n-1,k-1}_{\bm{\alpha},t}$,
    the code $D^{n,k}_{\bm{\alpha},t}$ we constructed is a longer non-GRS MDS code,
    and only requires the code dimension $k$ to satisfy $3\le k \le n-3$.
    \item The authors of \cite{Abdukhalikov, Liyang2} studied the extended codes of $C^{n,k}_{\bm{\alpha},t}$ and the conditions for them to be non-GRS MDS codes.
    From the construction method of $D^{n,k}_{\bm{\alpha},t}$,
    it follows that for these two classes of extended codes, longer non-GRS MDS codes with looser dimension constraints can be obtained via a construction analogous to that of $D^{n,k}_{\bm{\alpha},t}$.
    \item Assume that $6 \le 2k < n+1$, $1\le s \le k-1$, $\bm{\beta}=(\beta_1,\beta_2,...,\beta_n)$, and that $\{\beta_i\}$ are distinct elements of $\mathbb{F}_q$.
    Then the codes $D^{n,k}_{\bm{\alpha},t}$ and $C^{n,k}_{\bm{\beta},s}$ are not equivalent.
    This is because the punctured code of $D^{n,k}_{\bm{\alpha},t}$ at the last coordinate is a GRS code,
    whereas the punctured codes of $C^{n,k}_{\bm{\beta},s}$ at any coordinate are not GRS codes.
\end{enumerate}
\end{Remark}

\begin{Theorem}\label{NGRSTH}
    Assume that $3 \le k \le n-3$ and $1\le t \le k-1$. Then:
\begin{enumerate}
    \item If $\alpha_1,\alpha_2,...,\alpha_{n-1}$ are nonzero elements of $\mathbb{F}_q$,
    then the code $\MGRS_{n,k}(\bm{\alpha}, \bm{v}, \eta, t)$ is a GRS code if and only if $\eta = 0$.
    \item If there exists $\alpha_j=0$ where $1 \le j \le n-1$,
    then the code $\MGRS_{n,k}(\bm{\alpha}, \bm{v}, \eta, t)$ is not a GRS code.
\end{enumerate}
\begin{proof}
The code $C=\MGRS_{n,k}(\bm{\alpha}, \bm{v}, \eta, t)$ has a generator matrix $G = [\Diag(f_1(\alpha_1),f_2(\alpha_2),...,f_k(\alpha_k)) \mid A]\Diag(v_1,v_2,...,v_n)$,
\begin{align*}
    A = \begin{bmatrix}
        f_1(\alpha_{k+1}) & f_1(\alpha_{k+2}) & \cdots & f_1(\alpha_{n-1}) & \eta\sigma_{1,t} + \sigma_{1,0} \\
        f_2(\alpha_{k+1}) & f_2(\alpha_{k+2}) & \cdots & f_2(\alpha_{n-1}) & \eta\sigma_{2,t} + \sigma_{2,0} \\
        \vdots & \vdots & \cdots & \vdots & \vdots \\
         &  & \cdots &  &  \\
        f_{k}(\alpha_{k+1}) & f_{k}(\alpha_{k+2}) & \cdots & f_{k}(\alpha_{n-1}) & \eta\sigma_{k,t} + \sigma_{k,0}
        \end{bmatrix}.
\end{align*}

(1) When $\eta = 0$, $C$ is obviously a GRS code.
When $0 < t < k-1$, define 
$$
    m(x) = -a\eta x\sum_{j=0}^{t}\sigma_{P,j}x^{t-j} - b\eta\sum_{j=0}^{t}\sigma_{P,j}x^{t-j} - ax\sigma_{P,0} - b\sigma_{P,0} - 1.
$$
From the proof of Theorem \ref{SimNGRS}, if $\eta \ne 0$, the code $C$ is not a GRS code.

When $t=k-1$, assume $C$ is a GRS code.
Then it is an MDS code satisfying Corollary \ref{MDS1},
and all entries of matrix $A$ are nonzero.
Let
\[
    A_3 = \begin{bmatrix}
    \frac{1}{f_{h}(\alpha_x)} & \frac{1}{f_{h}(\alpha_y)} &  \frac{1}{f_{h}(0)+\eta}  \\
    \frac{1}{f_i(\alpha_x)} & \frac{1}{f_i(\alpha_y)} & \frac{1}{f_i(0)+\eta}  \\
    \frac{1}{f_j(\alpha_x)} & \frac{1}{f_j(\alpha_y)} & \frac{1}{f_j(0)+\eta}
    \end{bmatrix}, \quad 
    \begin{cases}
        a' = \frac{\alpha_x - \alpha_h}{P(\alpha_x)}, \quad b' = \frac{\alpha_y - \alpha_h}{P(\alpha_y)}, \quad c' = \frac{\alpha_h}{\alpha_h \eta - P(0)}, \\
        d' = \frac{\alpha_x - \alpha_i}{P(\alpha_x)}, \quad e' = \frac{\alpha_y - \alpha_i}{P(\alpha_y)}, \quad f' = \frac{\alpha_i}{\alpha_i \eta - P(0)}, \\
        g' = \frac{\alpha_x - \alpha_j}{P(\alpha_x)}, \quad h' = \frac{\alpha_y - \alpha_j}{P(\alpha_y)}, \quad i' = \frac{\alpha_j}{\alpha_j \eta - P(0)},
    \end{cases} 
\]
for $1\le h < i<j \le k$, $k+1 \le x<y\le n-1$ and $K = \frac{1}{P(\alpha_x)P(\alpha_y)}$.
We know $d'h' - e'g' = K(\alpha_x - \alpha_y)(\alpha_i - \alpha_j)$, $a'h' - g'b' = K(\alpha_x - \alpha_y)(\alpha_h - \alpha_j)$, $a'e' - b'd' = K(\alpha_x - \alpha_y)(\alpha_h - \alpha_i)$.
Then
\begin{align*}
\det(A_3) & = c'(d'h' - e'g') - f'(a'h' - g'b') + i'(a'e' - b'd') \\
    & = K(\alpha_x - \alpha_y) \cdot \left[
    \frac{\alpha_h (\alpha_i - \alpha_j)}{\alpha_h \eta - P(0)} -
    \frac{\alpha_i (\alpha_h - \alpha_j)}{\alpha_i \eta - P(0)} +
    \frac{\alpha_j (\alpha_h - \alpha_i)}{\alpha_j \eta - P(0)}\right] \\
    & = -\frac{P(0)\eta(\alpha_x - \alpha_y)(\alpha_h - \alpha_i)(\alpha_i - \alpha_j)(\alpha_j - \alpha_h)}{P(\alpha_x)P(\alpha_y)(\alpha_h \eta - P(0))(\alpha_i \eta - P(0))(\alpha_j \eta - P(0))}.
\end{align*}
By Lemma \ref{cauchy2} (3), $\det(A_3) = 0$, implying $\eta = 0$.

(2)
By contradiction: Without loss of generality, let $\alpha_{n-1} = 0$.
If $C$ were a GRS code, then by part (1) ($\alpha_{n-1} = 0$ does not affect the above derivation), $\eta$ would be $0$.
However, this makes the last two columns of $A$ identical, contradicting the MDS property.
Hence, $C$ is not a GRS code.

This completes the proof.
\end{proof}
\end{Theorem}

\subsection{Extended modified GRS codes}
\begin{Definition}\label{def2}
    For $1 \le k \le n$ and $1 \le t \le k-1$,  the {\em extended modified generalized Reed-Solomon} (extended modified GRS) code is defined as:
    $$\EMGRS_{n+1,k}(\bm{\alpha}, \bm{v}, \eta, t)=\left\{\left(v_1 f\left(\alpha_1\right), \ldots,v_{n-1}f\left(\alpha_{n-1}\right), v_n \left(f\left(0\right)+\eta f_t\right), v_{n+1}f_{k-1}\right) \mid f(x) \in \mathbb{F}_q[x]_{k}\right\},$$
    where $\eta \in \mathbb{F}_q$ and $f_t$ denotes the coefficient of $x^t$ in $f(x)$.
    Note that if $\eta=0$ and $\alpha_i \neq 0$, for all $i=1,...,n-1$, then $\EMGRS_{n+1,k}(\bm{\alpha}, \bm{v}, \eta, t)$ is the extended GRS code.
\end{Definition}

The code  $\EMGRS_{n+1,k}(\bm{\alpha}, \bm{v}, \eta, t)$ has the following generator matrix:
\begin{align} \label{EMGRMA}
    G = \begin{bmatrix}
    1 & 1 & \cdots & 1 & 1 & 0 \\
    \alpha_1 & \alpha_2 & \cdots & \alpha_{n-1} & 0 & 0 \\
    \vdots & \vdots & \cdots & \vdots & \vdots & \vdots \\
    \alpha_1^{t - 1} & \alpha_2^{t - 1} & \cdots & \alpha_{n-1}^{t - 1} & 0 & 0\\
    \alpha_1^{t} & \alpha_2^{t} & \cdots & \alpha_{n-1}^{t} & \eta & 0 \\
    \alpha_1^{t + 1} & \alpha_2^{t + 1} & \cdots & \alpha_{n-1}^{t + 1} & 0 & 0 \\
    \vdots & \vdots & \cdots & \vdots & \vdots & \vdots  \\
    \alpha_1^{k - 1} & \alpha_2^{k - 1} & \cdots & \alpha_{n-1}^{k - 1} & 0 & 1
    \end{bmatrix}\Diag(v_1,v_2,...,v_{n}, v_{n+1}).
\end{align}

\begin{Remark}
    The code $\EMGRS_{n+1,k}(\bm{\alpha}, \bm{v}, \eta, t)$ is an MDS or almost-MDS code.
\end{Remark}

\begin{Theorem}
    The code $\EMGRS_{n+1,k}(\bm{\alpha}, \bm{v}, \eta, t)$ ($t\neq k-1$) is MDS if and only if
for $m=k-2,k-1$, it holds that $\eta\pi_{t,m} \neq  (-1)^{m+1}\prod_{s=1}^{m} \alpha_{i_s}$
where $\{i_1, i_2, \dots, i_{m}\}$ is an arbitrary $m$-element subset of $\{1, 2, \dots, n-1\}$ and
$$\pi_{s,m} = (-1)^{m-s} \cdot \sum_{1 \leq j_1 < j_2 < \cdots < j_{m-s} \leq m} \alpha_{i_{j_1}}\alpha_{i_{j_2}}\cdots \alpha_{i_{j_{m-s}}} \ \text{for} \ s=0,...,m.$$
\begin{proof}
    Note that any $k$ columns among the first $n$ columns of $G$ (in (\ref{EMGRMA})) are linearly independent if and only if $\eta\pi_{t,k-1} \neq (-1)^{k}\prod_{s=1}^{k-1} \alpha_{i_s}$.
    The $(n+1)$-th column of $G$ is linearly independent of any $(k-1)$ columns among the first $n$ columns if and only if $ \eta\pi_{t,k-2} \neq (-1)^{k-1}\prod_{s=1}^{k-2} \alpha_{i_s} $.
    This completes the proof.
\end{proof}
\end{Theorem}

The following corollaries are easily derived from the above results.
\begin{Corollary}\label{extend+}
   The  code $\EMGRS_{n+1,k}(\bm{\alpha}, \bm{v}, \eta, 1)$ is MDS if and only if $\frac{1}{\eta} \neq \sum_{s=1}^{m}\frac{1}{\alpha_{i_s}}$,
   where $\{i_1, i_2, \dots, i_m\}$ is any $m$-element subset of $\{1, 2, \dots, n-1\}$ and $m \in \{k-2, k-1\}$.
\end{Corollary}

\begin{Corollary}
    If $\alpha_1,\alpha_2,\dots,\alpha_{n-1}$ are nonzero elements of $\mathbb{F}_q$,
then $\EMGRS_{n+1,k}(\bm{\alpha}, \bm{v}, \eta, k-1)$ is MDS if and only if $\MGRS_{n,k}(\bm{\alpha}, \bm{v}, \eta, k-1)$ is MDS.
\end{Corollary}

\begin{Corollary}\label{extendNGRS}
    If $\alpha_1,\alpha_2,\dots,\alpha_{n-1}$ are nonzero elements of $\mathbb{F}_q$,
$3 \le k\le n-4$ and $1\le t \le k-1$, then $\EMGRS_{n+1,k}(\bm{\alpha}, \bm{v}, \eta, t)$ is an extended GRS code if and only if $\eta = 0$.
\end{Corollary}

Corollary \ref{extendNGRS} also holds for $k=n-3$, and its proof can be derived similarly to that of Theorem \ref{NGRSTH}.
\begin{Remark}
    Assume that $6 \le 2k < n+1$, $1\le t \le k-1$, $\bm{\beta}=(\beta_1,\beta_2,...,\beta_n)$,
    and that $\{\beta_i\}$ are distinct elements of $\mathbb{F}_q$.
    \begin{enumerate}
        \item $\MGRS_{n,k}(\bm{\alpha}, \bm{v}, \eta, t)$ and $\EMGRS_{n,k}(\bm{\alpha}, \bm{v}, \eta, t)$ are not equivalent to twisted GRS codes $C^{n,k}_{\bm{\beta},1,s,\lambda}$ where $0\le s\le k-1$ (see \cite[Definition 1]{Beelen1}).
        \item $\MGRS_{n,k}(\bm{\alpha}, \bm{v}, \eta, t)$ and $\EMGRS_{n,k}(\bm{\alpha}, \bm{v}, \eta, t)$ are not equivalent to the codes $C^{n,k}_{\bm{\beta},s}$ where $1 \le s \le k-1$ (see \cite[Section VI]{Jin}), and are not equivalent to the codes associated with $C^{n,k}_{\bm{\beta},s}$ (see \cite[Definition 1]{Liyang2} and \cite[Page 6]{Abdukhalikov}).
    \end{enumerate}
    This is because the punctured code of $\MGRS_{n,k}(\bm{\alpha}, \bm{v}, \eta, t)$ (resp. $\EMGRS_{n,k}(\bm{\alpha}, \bm{v}, \eta, t)$) at the $n$-th (resp. $(n-1)$-th) coordinate is a GRS code,
whereas the punctured codes of the other codes mentioned herein at any coordinate position are not GRS codes.
\end{Remark}

\section{Relationships of non-GRS MDS codes}
From our constructions and previous research, the class of non-GRS MDS codes include
the (extended) modified GRS codes defined in Definitions \ref{def1} and \ref{def2},
the (extended) column-twisted GRS codes \cite{Liu},
the (row) twisted GRS codes \cite{Beelen1},
and the Roth-Lempel codes \cite{Roth}.
In this section, we analyze the interrelationships between these code classes.

\subsection{Roth-Lempel codes and column-twisted GRS codes}

This section elaborates on how Roth-Lempel codes and column-twisted GRS codes relate to the (extended) modified GRS codes given in this paper.
\begin{Definition}\cite{Roth}
Let $n$ and $k$ be integers such that $k \geq 3$ and $k + 3 \leq n \leq q + 2 $.
The $[n, k]$ {\em Roth-Lempel} code over $\mathbb{F}_q$ generated by the matrix
\begin{align}\label{RLMatrix}
    G = \begin{bmatrix}
        1 & 1 & \cdots & 1 & 0 & 0 \\
        a_1 & a_2 & \cdots & a_{n - 2} & 0 & 0 \\
        \vdots & \vdots & \cdots & \vdots & \vdots & \vdots \\
         &  & \cdots &  & 0 & 1 \\
        a_1^{k - 1} & a_2^{k - 1} & \cdots & a_{n - 2}^{k - 1} & 1 & \delta
        \end{bmatrix},
\end{align}
where $a_1,...,a_{n-2}$ are distinct elements of $\mathbb{F}_q$ and $\delta \in \mathbb{F}_q$.
\end{Definition}

\begin{Proposition}\label{PropRL}
    When $\delta \neq 0$, a Roth-Lempel code is equivalent to either a modified GRS code $\MGRS_{n,k}(\bm{\alpha}, \bm{v}, \eta, 1)$ or
an extended modified GRS code $\EMGRS_{n,k}(\bm{\alpha}, \bm{v}, \eta, 1)$.
\begin{proof}
    If $a_s \neq 0$ for all $1 \le s \le n-2$,
    then the generator matrix in (\ref{RLMatrix}) of a Roth-Lempel code can be transformed to
    \[
        G' = \begin{bmatrix}
            1 & 1 & \cdots & 1 & 1 & 1 \\
            \frac{1}{a_1} & \frac{1}{a_2} & \cdots & \frac{1}{a_{n - 2}} & 0 & \frac{1}{\delta}\\
            \vdots & \vdots & \cdots & \vdots & 0 & 0 \\
             \frac{1}{a_1^{k - 1}} & \frac{1}{a_2^{k - 1}} & \cdots & \frac{1}{a_{n - 2}^{k - 1}} & 0 & 0
            \end{bmatrix}\Diag(a_1^{k - 1}, a_2^{k - 1},...,a_{n-2}^{k - 1},1,\delta).
    \]
    Then the code is a modified GRS code $\MGRS_{n,k}(\bm{\alpha}, \bm{v}, \eta, 1)$,
    where $\alpha_i= a_i^{-1}, v_i=a_i^{k-1}$ for $i=1,...,n-2$, $\alpha_{n-1}=0, v_{n-1}=1, v_n=\delta$ and $\eta=\frac{1}{\delta}$.

    If there exists $a_s = 0$ for some $s$, then the generator matrix in (\ref{RLMatrix}) of a Roth-Lempel code can be transformed to
    \[
        G'' = \begin{bmatrix}
            1 & \cdots & 1 & 0 & 1 & \cdots & 1 & 1 & 1 \\
            \frac{1}{a_1} &  \cdots & \frac{1}{a_{s - 1}} & 0 & \frac{1}{a_{s+1}} & \cdots & \frac{1}{a_{n-2}} & 0 & \frac{1}{\delta}\\
            \vdots &  \cdots & \vdots & \vdots &  \vdots& \cdots & \vdots & 0 & 0 \\
            &  \cdots & & 0 &  & \cdots & &  &  \\
            \frac{1}{a_1^{k - 1}}  & \cdots & \frac{1}{a_{s - 1}^{k - 1}} & 1 & \frac{1}{a_{s + 1}^{k - 1}} & \cdots  & \frac{1}{a_{n - 2}^{k-1}} &  0 & 0
            \end{bmatrix}\Diag(a_1^{k - 1},...,a_{s-1}^{k - 1}, 1,a_{s+1}^{k - 1},..., a_{n-2}^{k - 1},1,\delta).
    \]
    Then the code is permutation-equivalent to an extended modified GRS code $\EMGRS_{n,k}(\bm{\alpha}, \bm{v}, \eta, 1)$.
\end{proof}
\end{Proposition}

\begin{Definition}\cite{Liu}
    For distinct $ a_1,a_2,...,a_{n-1}, b, c \in \mathbb{F}_q $,
the {\em column-twisted GRS} code has generator matrix
\[
    G_1(b, c) = \begin{pmatrix}
    1 & 1 & \cdots & 1 & 1 - \lambda \\
    a_1 & a_2 & \cdots & a_{n-1} & b - \lambda c \\
    \vdots & \vdots & \vdots & \vdots & \vdots \\
    a_1^{k-1} & a_2^{k-1} & \cdots & a_{n-1}^{k-1} & b^{k-1} - \lambda c^{k-1}
    \end{pmatrix}
\]
and the {\em extended column-twisted GRS} code has generator matrix
\[
    G_1(b, c, \infty) = \begin{pmatrix}
    1 & 1 & \cdots & 1 & 1 - \lambda & 0 \\
    a_1 & a_2 & \cdots & a_{n-1} & b - \lambda c & 0 \\
    \vdots & \vdots & \vdots & \vdots & \vdots & \vdots \\
    a_1^{k-2} & a_2^{k-2} & \cdots & a_{n-1}^{k-2} & b^{k-2} - \lambda c^{k-2} & 0 \\
    a_1^{k-1} & a_2^{k-1} & \cdots & a_{n-1}^{k-1} & b^{k-1} - \lambda c^{k-1} & 1
    \end{pmatrix},
\]
where $\lambda \in \mathbb{F}_q$.
\end{Definition}

\begin{Proposition}
\begin{enumerate}
    \item Column-twisted GRS codes are subclasses of modified GRS codes $\MGRS_{n,k}(\bm{\alpha}, \bm{v}, \eta, k-1)$.
    \item An extended column-twisted GRS code is permutation-equivalent to a modified GRS code $\MGRS_{n+1,k}(\bm{\alpha}, \bm{v}, \eta, k-1)$.
\end{enumerate}
\begin{proof}
    When $\lambda = 0$, the column-twisted GRS code reduces to a GRS code, and the proposition holds trivially.
Now consider the case $\lambda \neq 0$.

(1) We transform the generator matrix of a column-twisted GRS code to the following form:
    \begin{align*}
        G_1(b, c) \rightarrow &\begin{bmatrix}
            1 & 1 & \cdots & 1 & 1 - \lambda \\
            a_1-b & a_2-b & \cdots & a_{n-1}-b &  - \lambda (c-b) \\
            \vdots & \vdots & \vdots & \vdots & \vdots \\
            (a_1-b)^{k-1} & (a_2-b)^{k-1} & \cdots & (a_{n-1}-b)^{k-1} &  - \lambda (c-b)^{k-1}
            \end{bmatrix} \\
        \rightarrow &\begin{bmatrix}
            1 & 1 & \cdots & 1 & 1 \\
            \frac{1}{a_1-b} & \frac{1}{a_2-b} & \cdots & \frac{1}{a_{n-1}-b} & \frac{1}{c-b} \\
            \vdots & \vdots & \cdots & \vdots & \vdots \\
            \frac{1}{(a_1-b)^{k - 2}} & \frac{1}{(a_2-b)^{k - 2}} & \cdots & \frac{1}{(a_{n-1}-b)^{k - 2}} & \frac{1}{(c-b)^{k-2}} \\
            \frac{1}{(a_1-b)^{k - 1}} & \frac{1}{(a_2-b)^{k - 1}} & \cdots & \frac{1}{(a_{n-1}-b)^{k - 1}} & \frac{1}{(c-b)^{k-1}}-\frac{1}{\lambda(c-b)^{k-1}}
            \end{bmatrix}\\
            &\cdot\Diag((a_1-b)^{k-1},...,(a_{n-1}-b)^{k-1}, -\lambda(c-b)^{k-1})\\
        \rightarrow &\begin{bmatrix}
         1 & 1 & \cdots & 1 & 1 \\
    \alpha_1 & \alpha_2 & \cdots & \alpha_{n-1} & 0 \\
    \vdots & \vdots & \cdots & \vdots & \vdots \\
    \alpha_1^{k - 2} & \alpha_2^{k - 2} & \cdots & \alpha_{n-1}^{k - 2} & 0 \\
    \alpha_1^{k - 1} & \alpha_2^{k - 1} & \cdots & \alpha_{n-1}^{k - 1} & \eta
    \end{bmatrix}\Diag(v_1,v_2,...,v_{n}).
\end{align*}

This is the generator matrix of a modified GRS code $\MGRS_{n,k}(\bm{\alpha}, \bm{v}, \eta, k-1)$,
where $\alpha_i = \frac{1}{a_i-b}-\frac{1}{c-b}, v_i=(a_i-b)^{k-1}$, for $i = 1,...,n-1$, $v_n=-\lambda(c-b)^{k-1}$ and $\eta = -\frac{1}{\lambda (c-b)^{k-1}}$.
Since all $\alpha_i$ here are nonzero, column-twisted GRS codes are subclasses of modified GRS codes $\MGRS_{n,k}(\bm{\alpha}, \bm{v}, \eta, k-1)$.

(2) We transform the generator matrix of an extended column-twisted GRS code to the following form:
\begin{align*}
    G_1(b, c, \infty) \rightarrow &\begin{bmatrix}
        1 & 1 & \cdots & 1 & 1 - \lambda & 0 \\
        a_1-b & a_2-b & \cdots & a_{n-1}-b &  - \lambda (c-b) & 0 \\
        \vdots & \vdots & \vdots & \vdots & \vdots & \vdots \\
        (a_1-b)^{k-1} & (a_2-b)^{k-1} & \cdots & (a_{n-1}-b)^{k-1} &  - \lambda (c-b)^{k-1} & 1
        \end{bmatrix} \\
        \rightarrow &\begin{bmatrix}
        1 & 1 & \cdots & 1 & 1 & 1 \\
        \frac{1}{a_1-b} & \frac{1}{a_2-b} & \cdots & \frac{1}{a_{n-1}-b} & \frac{1}{c-b} & 0 \\
        \vdots & \vdots & \cdots & \vdots & \vdots & \vdots \\
        \frac{1}{(a_1-b)^{k - 2}} & \frac{1}{(a_2-b)^{k - 2}} & \cdots & \frac{1}{(a_{n-1}-b)^{k - 2}} & \frac{1}{(c-b)^{k-2}} & 0 \\
        \frac{1}{(a_1-b)^{k - 1}} & \frac{1}{(a_2-b)^{k - 1}} & \cdots & \frac{1}{(a_{n-1}-b)^{k - 1}} & \frac{1}{(c-b)^{k-1}}-\frac{1}{\lambda(c-b)^{k-1}} & 0
        \end{bmatrix}\\
        &\cdot \Diag((a_1-b)^{k-1},...,(a_{n-1}-b)^{k-1}, -\lambda(c-b)^{k-1}, 1)\\
        \rightarrow &\begin{bmatrix}
        1 & 1 & \cdots & 1 & 1 & 1\\
    \alpha_1 & \alpha_2 & \cdots & \alpha_{n-1} & 0 & \alpha_n \\
    \vdots & \vdots & \cdots & \vdots & \vdots & \vdots \\
    &  & \cdots &  & & \\
    \alpha_1^{k - 2} & \alpha_2^{k - 2} & \cdots & \alpha_{n-1}^{k - 2} & 0 &  \alpha_{n}^{k - 2} \\
    \alpha_1^{k - 1} & \alpha_2^{k - 1} & \cdots & \alpha_{n-1}^{k - 1} & \eta & \alpha_{n}^{k - 1}
    \end{bmatrix}\Diag(v_1,v_2,...,v_{n},v_{n+1}),
\end{align*}
where $\alpha_i = \frac{1}{a_i-b}-\frac{1}{c-b}, v_i=(a_i-b)^{k-1}$, for $i = 1,...,n-1$, $\alpha_n=-\frac{1}{c-b}$, $v_n=-\lambda(c-b)^{k-1}$, $v_{n+1}=1$ and $\eta = -\frac{1}{\lambda (c-b)^{k-1}}$.
Thus, an extended column-twisted GRS code is permutation-equivalent to a modified GRS code $\MGRS_{n+1,k}(\bm{\alpha}, \bm{v}, \eta, k-1)$.
\end{proof}
\end{Proposition}

\subsection{Twisted GRS codes}
Compared with the definition of the original twisted GRS codes in \cite{Beelen1},
what we have consider here is actually two special types of twisted GRS codes.
Moreover, we present the relationship between these two special types of twisted GRS codes and other non-GRS MDS codes.

\begin{Definition}\cite{Beelen1}\label{TGRS}
For $1 \le k \le n$. {\em Twisted generalized Reed-Solomon} (twisted GRS) $\TGRS_{n+1,k}(\bm{\alpha}, \bm{v}, \lambda, 0)$ has a generator matrix
\[
    G_1 =\begin{pmatrix}
    1+\lambda\alpha_1^k & \cdots & 1+\lambda\alpha_n^k & 1 \\
    \alpha_1 & \cdots & \alpha_n & 0 \\
    \vdots & & \vdots & \vdots \\
    \alpha_1^{k-2} & \cdots & \alpha_n^{k-2} & 0 \\
    \alpha_1^{k-1} & \cdots & \alpha_n^{k-1} & 0
    \end{pmatrix}\Diag(v_1,v_2,...,v_n,v_{n+1}),
\]
where $\lambda \in \mathbb{F}_q^*$.
{\em Twisted GRS code} $\TGRS_{n+1,k}(\bm{\alpha}, \bm{v}, \lambda, k-1)$ has a generator matrix
\[
    G_2 =\begin{pmatrix}
        1 & \cdots & 1 & 0\\
        \alpha_1 & \cdots & \alpha_n & 0 \\
        \vdots & & \vdots & \vdots\\
        \alpha_1^{k-2} & \cdots & \alpha_n^{k-2} & 0 \\
        \alpha_1^{k-1}+\lambda\alpha_1^k & \cdots & \alpha_n^{k-1}+\lambda\alpha_n^k & 1
    \end{pmatrix}\Diag(v_1,v_2,...,v_n,v_{n+1}),
\]
where $\lambda \in \mathbb{F}_q^*$.
\end{Definition}

\begin{Proposition}\label{Propcon1}
If $\alpha_1,\alpha_2,...,\alpha_{n}$ are nonzero elements of $\mathbb{F}_q$,
then the twisted GRS code $\TGRS_{n+1,k}(\bm{\alpha}, \bm{v}, \lambda, 0)$ is the dual of the modified GRS code $\MGRS_{n+1,k}(\bm{\alpha}, \bm{w}, \eta, k-1)$,
which has a parity-check matrix as
\begin{align}\label{Check1}
    H_1 = \begin{pmatrix}
        1& \cdots &  1 & 1 \\
        \alpha_1& \cdots  & \alpha_n & 0\\
        \vdots & \ddots & \ddots & \vdots \\
        \alpha_1^{n-k-1}& \cdots &   \alpha_n^{n-k-1} & 0 \\
        \alpha_1^{n-k}& \cdots &   \alpha_n^{n-k} & \eta
    \end{pmatrix}\Diag(w_1,w_2,...,w_n,w_{n+1}),
\end{align}
where $u_i^{-1}=\prod_{j=1,j\ne i}^n(\alpha_i-\alpha_j), w_i=u_i/(\alpha_iv_i), (1\le i \le n)$, $w_{n+1}=-(\sum_{i=1}^nu_i\alpha_i^{-1})/v_{n+1}$, $\eta=\lambda(\sum_{i=1}^nu_i\alpha_i^{n-1})/(\sum_{j=1}^nu_j\alpha_j^{-1})$ and $\bm{w}=(w_1,w_2,...,w_{n+1})$.

\begin{proof}
We know $\langle (\alpha_1^j,...,\alpha_n^j), (u_1,...,u_n) \rangle = 0$ for $0 \le j \le n-2$,  where $u_i^{-1}=\prod_{j=1,j\ne i}^n(\alpha_i-\alpha_j),\ (1\le i \le n)$, by Proposition \ref{GRSDual}.

Now, we prove that the code
$\TGRS_{n+1,k}(\bm{\alpha}, \bm{v}, \lambda, 0)$ has a parity check matrix as (\ref{Check1}).
We only need to show that $\langle H_1[1], G_1[1] \rangle = 0$ and $\langle H_1[n-k+1], G_1[1] \rangle = 0$, where $H_1[i]$ denotes the $i$-th row of the matrix $H_1$.
\begin{align*}
    \langle H_1[1], G_1[1] \rangle & = \sum_{i=1}^n(1+\lambda\alpha_i^k)v_i \cdot w_i + v_{n+1}w_{n+1} \\
    & = \sum_{i=1}^n(\alpha_i^{-1}+\lambda\alpha_i^{k-1})u_i + v_{n+1}\left(-\left(\sum_{i=1}^nu_i\alpha_i^{-1}\right)/v_{n+1}\right) = 0. \\
    \langle H_1[n-k+1], G_1[1] \rangle & = \sum_{i=1}^n(1+\lambda\alpha_i^k)v_i \cdot \alpha_i^{n-k}w_i + \eta v_{n+1}w_{n+1} \\
    & = \sum_{i=1}^n(\alpha_i^{n-k-1}+\lambda\alpha_i^{n-1})u_i + \eta v_{n+1}\left(-\left(\sum_{i=1}^nu_i\alpha_i^{-1}\right)/v_{n+1}\right) = 0.
\end{align*}
The proof is completed.
\end{proof}
\end{Proposition}

\begin{Corollary}
    Assume $3 \le k \le n-2$. Then:
\begin{enumerate}
    \item If $\alpha_1,\alpha_2,...,\alpha_{n-1}$ are nonzero elements of $\mathbb{F}_q$,
    then the code $\TGRS_{n+1,k}(\bm{\alpha}, \bm{v}, \lambda, 0)$ is GRS code if and only if $\lambda = 0$.
    \item If there exists some $\alpha_j=0$, where $1 \le j \le n$,
    then the code $\TGRS_{n+1,k}(\bm{\alpha}, \bm{v}, \lambda, 0)$ is not a GRS code.
\end{enumerate}
\begin{proof}
    (1) The dual code of a GRS code is also a GRS code. Therefore, the conclusion can be directly obtained from Theorem \ref{NGRSTH} and Proposition \ref{Propcon1}.

    (2) Without loss of generality, assume $\alpha_n = 0$.
    In this case, the last two columns of the code $\TGRS_{n+1,k}(\bm{\alpha}, \bm{v}, \lambda, 0)$ are identical. So it cannot be an MDS code.
\end{proof}
\end{Corollary}

A similar proof for the following result can be found in \cite{Zhu}.
\begin{Proposition}\label{Propcon2}
The dual code of the twisted GRS code $\TGRS_{n+1,k}(\bm{\alpha}, \bm{v}, \lambda, k-1)$, which is equivalent to a punctured code of a Roth-Lempel code in the penultimate column,
has a generator matrix given by
\begin{align}\label{Check2}
    H_2 = \begin{pmatrix}
        1& \cdots  & 1 & 0 \\
        \vdots & \ddots &  & \vdots \\
        \alpha_1^{n-k-2}& \cdots  & \alpha_n^{n-k-2} & 0 \\
        \alpha_1^{n-k-1}& \cdots  & \alpha_n^{n-k-1} & 1 \\
        \alpha_1^{n-k}& \cdots  & \alpha_n^{n-k} & \delta
    \end{pmatrix}\Diag(w_1,w_2,...,w_n,w_{n+1}),
\end{align}
where $u_i^{-1}=\prod_{j=1,j\ne i}^n(\alpha_i-\alpha_j), w_i=u_i/v_i$, for $1\le i \le n$, $w_{n+1}=-\lambda(\sum_{i=1}^nu_i\alpha_i^{n-1})/v_{n+1}$, $\delta=\sum_{i=1}^nu_i(\alpha_i^{n-1}+\lambda \alpha_i^{n})/(\lambda(\sum_{i=1}^nu_i\alpha_i^{n-1}))$ and $\bm{w}=(w_1,w_2,...,w_{n+1})$.

\begin{proof}
    By Proposition \ref{GRSDual}, we have $\langle (\alpha_1^j,...,\alpha_n^j), (u_1,...,u_n) \rangle = 0$ for $0 \le j \le n-2$, where $u_i^{-1}=\prod_{j=1,j\ne i}^n(\alpha_i-\alpha_j)$, for $1\le i \le n$.

We now prove that the code
$\TGRS_{n+1,k}(\bm{\alpha}, \bm{v}, \lambda, k-1)$ has the parity-check matrix given in (\ref{Check2}).
We only need to show that $\langle H_2[n-k], G_2[k] \rangle = 0$ and $\langle H_2[n-k+1], G_2[k] \rangle = 0$.
\begin{align*}
    \langle H_2[n-k], G_2[k] \rangle & = \sum_{i=1}^nw_i\alpha_i^{n-k-1} \cdot (\alpha_i^{k-1}+\lambda \alpha_i^k)v_i + w_{n+1}v_{n+1} \\
    & = \sum_{i=1}^nu_i(\alpha_i^{n-2} + \lambda \alpha_i^{n-1}) + \left(-\lambda(\sum_{i=1}^nu_i\alpha_i^{n-1})/v_{n+1} \right) v_{n+1} = 0. \\
    \langle H_2[n-k+1], G_2[k] \rangle & = \sum_{i=1}^nw_i\alpha_i^{n-k} \cdot (\alpha_i^{k-1}+\lambda \alpha_i^k)v_i + \delta w_{n+1}v_{n+1} \\
    & = \sum_{i=1}^nu_i(\alpha_i^{n-1} + \lambda \alpha_i^{n}) + \delta\left(-\lambda\sum_{i=1}^nu_i\alpha_i^{n-1}\right) = 0.
\end{align*}
This completes the proof.
\end{proof}
\end{Proposition}

\begin{Lemma}\label{punRLng}
    Assume $3 \le k \le n - 3$. A punctured code of a Roth-Lempel code in the penultimate column is not a GRS code,
whose generator matrix is given by
    \begin{align*}
        G = \begin{bmatrix}
            1 & 1 & \cdots & 1 & 0 \\
            \alpha_1 & \alpha_2 & \cdots & \alpha_{n - 1} & 0 \\
            \vdots & \vdots & \cdots & \vdots & \vdots  \\
             &  & \cdots &  &  1 \\
             \alpha_1^{k - 1} & \alpha_2^{k - 1} & \cdots & \alpha_{n - 1}^{k - 1} & \delta
            \end{bmatrix},
    \end{align*}
where $\delta \in \mathbb{F}_q$.

\begin{proof}
The code $C=\MGRS_{n,k}(\bm{\alpha}, \bm{v}, \eta, k-1)$ has a generator matrix $G = [\Diag(f_1(\alpha_1),f_2(\alpha_2),...,f_k(\alpha_k)) \mid A]\Diag(v_1,v_2,...,v_n)$, where
\begin{align*}
    A = \begin{bmatrix}
        f_1(\alpha_{k+1}) & f_1(\alpha_{k+2}) & \cdots & f_1(\alpha_{n-1}) & K + \alpha_1 \\
        f_2(\alpha_{k+1}) & f_2(\alpha_{k+2}) & \cdots & f_2(\alpha_{n-1}) & K + \alpha_2 \\
        \vdots & \vdots & \cdots & \vdots & \vdots \\
            &  & \cdots &  &  \\
        f_{k}(\alpha_{k+1}) & f_{k}(\alpha_{k+2}) & \cdots & f_{k}(\alpha_{n-1}) & K + \alpha_k
        \end{bmatrix}.
\end{align*}
where $f_i(x)=\frac{\prod_{j=1}^k(x-\alpha_j)}{x-\alpha_i}$, for $i= 1,...,k$, and $K= \delta - \sum_{i=1}^k\alpha_i$.

If any entry of matrix $A$ is zero, then the code $C$ is neither MDS nor GRS.
Therefore, we assume that all entries of matrix $A$ are nonzero.
Let
    \[
        A_3 = \begin{bmatrix}
        \frac{1}{f_h(\alpha_x)} & \frac{1}{f_h(\alpha_y)} & \frac{1}{K+\alpha_h}  \\
        \frac{1}{f_i(\alpha_x)} & \frac{1}{f_i(\alpha_y)} & \frac{1}{K+\alpha_i}  \\
        \frac{1}{f_j(\alpha_x)} & \frac{1}{f_j(\alpha_y)} & \frac{1}{K+\alpha_j}
        \end{bmatrix}, \quad \begin{cases}
            a' = \frac{\alpha_x - \alpha_h}{P(\alpha_x)}, \quad b' = \frac{\alpha_y - \alpha_h}{P(\alpha_y)}, \quad c' = \frac{1}{K + \alpha_h }, \\
            d' = \frac{\alpha_x - \alpha_i}{P(\alpha_x)}, \quad e' = \frac{\alpha_y - \alpha_i}{P(\alpha_y)}, \quad f' = \frac{1}{K + \alpha_i }, \\
            g' = \frac{\alpha_x - \alpha_j}{P(\alpha_x)}, \quad h' = \frac{\alpha_y - \alpha_j}{P(\alpha_y)}, \quad i' = \frac{1}{K + \alpha_j },
            \end{cases}
    \]
for any $1\le h < i<j \le k$ and $k+1 \le x<y\le n-1$ where $P(x)= \prod_{j=1}^k(x-\alpha_j)$. We have
\begin{align*}
    \det(A_3)= &c'(d'h' - e'g') - f'(a'h' - g'b') + i'(a'e' - b'd') \\
        = & \frac{(\alpha_x - \alpha_y)}{P(\alpha_x)P(\alpha_y)} \left( \frac{\alpha_i - \alpha_j}{K + \alpha_h} - \frac{\alpha_h - \alpha_j}{K + \alpha_i} + \frac{\alpha_h - \alpha_i}{K + \alpha_j} \right) \\
        = & -\frac{(\alpha_x - \alpha_y)}{P(\alpha_x)} \frac{(\alpha_i-\alpha_j)(\alpha_j-\alpha_h)(\alpha_h-\alpha_i)}{(K + \alpha_h)(K + \alpha_i)(K + \alpha_j)}.
\end{align*}
Thus $\det(A_3) \neq 0$. By Lemma \ref{cauchy2} (3), the code $C$ is not a GRS code.
\end{proof}
\end{Lemma}

Although the following result was already proven in \cite{Roth},
we provide here a more concise proof based on the Cauchy matrix.
\begin{Corollary}
    Assume $3 \le k \le n - 3$. An $[n, k]$ Roth-Lempel code is not a GRS code.
\begin{proof}
    In fact, it suffices to slightly modify the proof of Lemma \ref{punRLng}: replacing the penultimate column of matrix $G$ in Lemma \ref{punRLng} with $(0, \ldots, 0, 1)^T$ yields a generator matrix of the $[n, k]$ Roth-Lempel code.
    In the proof of Lemma 4.1, correspondingly, replace the penultimate column of matrix $A$ with an all-ones vector, and let $b', e', h'$ be replaced by $b'=1$, $e'=1$, $h'=1$ respectively. In this case, we have
\begin{align*}
    \det(A_3) &= c'(d'h' - e'g') - f'(a'h' - g'b') + i'(a'e' - b'd') \\
    &= -\frac{1}{P(\alpha_x)}\left( \frac{\alpha_i - \alpha_j}{K + \alpha_h} - \frac{\alpha_h - \alpha_j}{K + \alpha_i} + \frac{\alpha_h - \alpha_i}{K + \alpha_j}\right) \\
    &= \frac{1}{P(\alpha_x)} \frac{(\alpha_i-\alpha_j)(\alpha_j-\alpha_h)(\alpha_h-\alpha_i)}{(K + \alpha_h)(K + \alpha_i)(K + \alpha_j)}.
\end{align*}
Thus $\det(A_3) \neq 0$. By Lemma \ref{cauchy2} (3), the $[n, k]$ Roth-Lempel code is not a GRS code.
\end{proof}
\end{Corollary}

The following result is already contained in [Theorem 5]\cite{Liyang},
and we can derive it in a more concise manner using the conclusions proven above.
\begin{Corollary}\label{NGR}
    If $3 \le k \le n-2$, twisted GRS codes $\TGRS_{n+1,k}(\bm{\alpha}, \bm{v}, \lambda, k-1)$ are non-GRS codes.
\begin{proof}
    It follows directly from Proposition \ref{Propcon2} and Lemma \ref{punRLng}.
\end{proof}
\end{Corollary}

It follows from Propositions \ref{PropRL}, \ref{Propcon1}, and \ref{Propcon2} that decoding modified GRS codes only requires adjustments to the existing decoding algorithms \cite{Sui3, Sun1, Wang}.

\begin{Remark}
It is noteworthy that other types of codes can be constructed by shortening the extended modified GRS code, as specified below:
\begin{enumerate}
    \item The shortened code of the modified GRS code at position $n$ is a twisted GRS code;
    \item If $\alpha_{n-1}=0$, the shortened code of the modified GRS code at the position set $\{n-1, n\}$ is a subcode of the GRS code (see \cite{Jin} for related research);
    \item If $\alpha_{n-1}=0$, the shortened code of the extended modified GRS code at the position set $\{n-1, n\}$ is a subcode of the extended GRS code (see \cite{Abdukhalikov, Liyang2} for related research).
\end{enumerate}
\end{Remark}

\section{Code length of non-GRS (non-elliptic curve) MDS codes}
Now let us discuss the code lengths of modified GRS, twisted GRS and Roth-Lempel codes, which are non-GRS MDS codes.
Additionally, we will present a range of constructions for non-GRS MDS codes.
Let $\mathbb{F}_q = \mathbb{F}_p(w)$, where $q=p^s$, $p$ is a prime, and $w$ is a primitive element.
Unless otherwise specified, $\bm{v}$ is the all-one vector with entries from $\mathbb{F}_q$.

\begin{Definition}
    Let $G$ be a proper subgroup of $(\mathbb{F}_q^*, \cdot)$, $\alpha_i \in G \cup\{0\}$ for $1 \le i \le n$, and $(-1)^k\eta \in \mathbb{F}_q^*\setminus G$.
Then, we call $\MGRS_{n,k}(\bm{\alpha}, \bm{v}, \eta, k-1)$ a $(*)$-modified GRS code.
\end{Definition}

\begin{Remark}
    Any $(*)$-modified GRS code is MDS.
\end{Remark}

\begin{Definition}
    Let $V$ be a proper subgroup of $(\mathbb{F}_q, +)$, $\eta^{-1} \in \mathbb{F}_q\setminus V$, and $\alpha_i^{-1} \in V\setminus\{0\} \cup \{\infty\} $ for $1 \le i \le n$.
Then, we call $\MGRS_{n,k}(\bm{\alpha}, \bm{v}, \eta, 1)$ a $(+)$-modified GRS code.
\end{Definition}

\begin{Remark}
    Any $(+)$-modified GRS code is MDS.
\end{Remark}

\begin{Theorem}\label{OddLength}
    If $p$ is an odd prime, then the $(*)$-modified GRS codes can yield non-GRS MDS codes with code length $\frac{q+3}{2}$.
\begin{proof}
    We know $w$ is a generator of $\mathbb{F}_q^*$; then $w^2$ is a generator of $G$,
where $G$ is a subgroup of $\mathbb{F}_q^*$ with index $2$. Let $\alpha_i=(w^2)^i$, for $i = 1,2,...,n-2$, and $\alpha_{n-1} = 0$, where $n-2=\frac{q-1}{2}$.
Take $\eta'$ to be any element of $\mathbb{F}_q^*\setminus G$ and let $\eta = (-1)^k\eta'$; thus $(-1)^k\eta = \eta' \in \mathbb{F}_q^*\setminus G$.
We have $\prod_{t=1}^{k-1}\alpha_{i_t}$ belongs to $G$ or equal to zero, and $(-1)^{k} \eta \neq \prod_{t=1}^{k-1}\alpha_{i_t}$ for any $(k-1)$-element subset $\{i_1, i_2, \dots, i_{k-1}\}$ of $\{1, 2, \dots, n-1\}$.
Then, by Corollary \ref{MDS1} and Theorem \ref{NGRSTH} (2), the $(*)$-modified GRS code $\MGRS_{n,k}(\bm{\alpha}, \bm{v}, \eta, k-1)$ is a non-GRS MDS code with length $\frac{q+3}{2}$.
\end{proof}
\end{Theorem}

In fact, when $k = 3$ and $p$ is odd, we can construct a class of non-GRS MDS codes over $\mathbb{F}_q$ ($q=p^s, s > 1$) with their lengths reaching $\frac{q+5}{2}$. 
The results are as follows.

\begin{Theorem}\label{OddLength3}
    If $p$ is an odd prime, and $k = 3$ or $k=\frac{q-1}{2}$, then 
the $\MGRS_{n,k}(\bm{\alpha}, \bm{v}, \eta, k-1)$ codes can yield non-GRS MDS codes with code length $\frac{q+5}{2}$.
    \begin{proof}
        Let $\alpha_i = w^i$ for $i = 1,...,n-3$, $\alpha_{n-2}=1, \alpha_{n-1} = 0$ and $\eta = -1$, where $n-3 = \frac{q-1}{2}$.
    We have $\alpha_{i_1}\alpha_{i_2} \neq 1$, i.e., $\alpha_{i_1}\alpha_{i_2} \neq (-1)^3\eta$ for any $2$-element subset $\{i_1,i_2\}$ of $\{1,2,...,n-1\}$.
    Then, by Corollary \ref{MDS1} and Theorem \ref{NGRSTH} (2), the $(*)$-modified GRS code $\MGRS_{n,k}(\bm{\alpha}, \bm{v}, \eta, k-1)$ is a non-GRS MDS code with length $\frac{q+5}{2}$.
    Furthermore, its dual code is a non-GRS MDS code with parameters $[\frac{q+5}{2}, \frac{q-1}{2}, 4]$.
\end{proof}
\end{Theorem}

\begin{Example}
    Let $\mathbb{F}_{11} = \{0,1,2,...,10\}$, we know $w=2$ is a primitive element of $\mathbb{F}_{11}$.
    Let $\alpha_1 = w=2$, $\alpha_2 = w^2=4$, $\alpha_3 = w^3=8$, $\alpha_4 = w^4=5$, $\alpha_5 = w^5=10$, $\alpha_6 = 1$, $\alpha_7=0$ and $\eta = -1$.
    Then the generator matrix of the code $C = \MGRS_{8,3}(\bm{\alpha}, \bm{v}, \eta, 2$) is given by:
    \begin{align*}
        G = \begin{bmatrix}
        1 & 1 & 1 & 1 & 1 & 1 & 1 & 1 \\
        \alpha_1 & \alpha_2 & \alpha_3 & \alpha_4 & \alpha_5 & \alpha_6 & \alpha_7 & 0 \\
        \alpha_1^2 & \alpha_2^2 & \alpha_3^2 & \alpha_4^2 & \alpha_5^2 & \alpha_6^2 & \alpha_7^2 & \eta \\
        \end{bmatrix} = \begin{bmatrix}
            1 & 1 & 1 & 1 & 1 & 1 & 1 & 1 \\
            2 & 4 & 8 & 5 & 10 & 1 & 0 & 0 \\
            4 & 5 & 9 & 3 & 1 & 1 & 0 & -1 \\
        \end{bmatrix}.
    \end{align*}
Via Magma \cite{Bosma}, the code $C$ is easily verified to be an MDS code with parameters $[8,3,6]$.
It is further determined to be a non-GRS code using Algorithm 2.
\end{Example}

\begin{Lemma}\label{lenlem}
    If $p=2$, then the $(+)$-modified GRS codes can yield non-GRS MDS codes with code length $\frac{q+2}{2}$.
\begin{proof}
    We may assume that $q = 2^s$, where $s > 1$.
Recall that $\mathbb{F}_q$ is an $s$-dimensional linear space over $\mathbb{F}_2$,
and $1, w, \dots, w^{s-1}$ form a basis for this space.
Let $V$ be the vector space over $ \mathbb{F}_2$ spanned by $1, w, \dots, w^{s-2}$.
Denote the nonzero elements of $V$ as $\beta_1, \beta_2, \dots, \beta_{n-2}$,
where $n - 2 = \frac{q}{2} - 1$.
Let $ \alpha_i = \frac{1}{\beta_i} $ for $ i = 1, \dots, n-2 $, $ \alpha_{n-1} = 0 $, and $ \eta = \frac{1}{w^{s-1}} $.
For any $ (k-1) $-element subset $ \{i_1, i_2, \dots, i_{k-1}\} $ of $ \{1, 2, \dots, n-1\} $,
it holds that $ \sum_{t=1}^{k-1}\frac{1}{\alpha_{i_t}} $ either belongs to $ V $ or is equal to $ \infty $,
and $ \frac{1}{\eta} \neq \sum_{t=1}^{k-1}\frac{1}{\alpha_{i_t}} $.
Then, by Corollary \ref{MDS2} and Theorem \ref{NGRSTH} (2), the $(+)$-modified GRS code $\MGRS_{n,k}(\bm{\alpha}, \bm{v}, \eta, 1)$ is a non-GRS MDS code with length $\frac{q+2}{2}$.
\end{proof}
\end{Lemma}

According to the definition of the extended modified GRS code in Definition \ref{def2},
the definitions of the extended $(+)$-modified GRS code and the extended $(*)$-modified GRS code can be obtained similarly, which will not be repeated here.

\begin{Corollary}\label{core+}
    Any extended $(+)$-modified GRS code is MDS.
    \begin{proof}
        This result follows directly from Corollary \ref{extend+}.
    \end{proof}
\end{Corollary}

It is obvious that both the shortened and punctured codes of a GRS code at any coordinate position are GRS codes.
This implies that the (column) extended code of a non-GRS code cannot be a GRS code;
therefore, we have the following conclusion:

\begin{Theorem}\label{leng+}
    If $p=2$, then extended $(+)$-modified GRS codes can yield non-GRS MDS codes with code length $\frac{q+4}{2}$.
\begin{proof}
    This can be directly obtained from Corollary \ref{core+} and the proof of Lemma \ref{lenlem}.
\end{proof}
\end{Theorem}

In fact, when $k = 4$ and $p = 2$, we can construct a class of non-GRS MDS codes over $\mathbb{F}_q$ ($q=p^s, s > 1$) with their lengths reaching $\frac{q+6}{2}$. 
The results are as follows.

\begin{Theorem}\label{leng+4}
    If $p=2$, then when $k = 4$ or $k=\frac{q-2}{2}$,
the $\MGRS_{n,k}(\bm{\alpha}, \bm{v}, \eta, 1)$ codes can yield non-GRS MDS codes with code length $\frac{q+6}{2}$.
\begin{proof}
    We may assume that $ q = 2^s $, where $s > 2$.
As we know, $ \mathbb{F}_q $ is an $ s $-dimensional linear space over $ \mathbb{F}_p $,
and $1, w, \dots, w^{s-1}$ form a basis for this space.
Let $V$ be the vector space over $ \mathbb{F}_p $ spanned by $ 1, w, \dots, w^{s-2} $.

Denote the elements of $w^{s-1}+V $ as $\{\beta_1, \beta_2, \dots, \beta_{n-3} \}$, where $n - 3 = \frac{q}{2}$.
Let $ \alpha_i = \frac{1}{\beta_i} $ for $i = 1, \dots, n-3$, $\alpha_{n-2} = 0$, $\alpha_{n-1}=1$, and $\eta = 1$.
For any $3$-element subset $ \{i_1, i_2,i_3\} $ of $ \{1, 2, \dots, n-2\} $,
it holds that $ \sum_{t=1}^3\frac{1}{\alpha_{i_t}} $ either belongs to $w^{s-1}+ V$ or is equal to $ \infty $,
and $ \frac{1}{\eta} \neq \sum_{t=1}^3\frac{1}{\alpha_{i_t}} $.
For any $2$-element subset $ \{i_1, i_2\}$ of $\{1, 2, \dots, n-2\}$, we have $\frac{1}{\alpha_{i_1}} + \frac{1}{\alpha_{i_2}} \neq 0$ and
$\frac{1}{\alpha_{i_1}} + \frac{1}{\alpha_{i_2}} + \frac{1}{\alpha_{n-1}} \neq \frac{1}{\eta}$.
Then, by Corollary \ref{MDS2} and Theorem \ref{NGRSTH} (2), the code $\MGRS_{n,4}(\bm{\alpha}, \bm{v}, \eta, 1)$ is a non-GRS MDS code with length $\frac{q+6}{2}$.
And its dual code is a non-GRS MDS code with parameters $[\frac{q+6}{2}, \frac{q-2}{2}, 5]$.
\end{proof}
\end{Theorem}

\begin{Example}
    Let $\mathbb{F}_8 = \mathbb{F}_2(w)$ (where $w$ satisfies $w^3 = w + 1$) and let $V$ be the vector space over $\mathbb{F}_2$ spanned by $\{1, w\}$; it follows that $w^2 + V = \{w^2, w^4, w^5, w^6\}$.
    Set $\alpha_1 = \frac{1}{w^2} = w^5$, $\alpha_2 = \frac{1}{w^4} = w^3$, $\alpha_3 = \frac{1}{w^5} = w^2$, $\alpha_4 = \frac{1}{w^6} = w$, $\alpha_5 = 0$, $\alpha_6 = 1$, $\eta = 1$.
    The generator matrix of the code $C = \MGRS_{7,4}(\bm{\alpha}, \bm{v}, \eta, 1)$ is given by:
    \begin{align*}
        G = \begin{bmatrix}
        1 & 1 & 1 & 1 & 1 & 1 & 1 \\
        \alpha_1 & \alpha_2 & \alpha_3 & \alpha_4 & \alpha_5 & \alpha_6 & \eta \\
        \alpha_1^2 & \alpha_2^2 & \alpha_3^2 & \alpha_4^2 & \alpha_5^2 & \alpha_6^2 & 0 \\
        \alpha_1^3 & \alpha_2^3 & \alpha_3^3 & \alpha_4^3 & \alpha_5^3 & \alpha_6^3 & 0
        \end{bmatrix} = \begin{bmatrix}
            1 & 1 & 1 & 1 & 1 & 1 & 1 \\
            w^5 & w^3 & w^2 & w & 0 & 1 & 1 \\
            w^3 & w^6 & w^4 & w^2 & 0 & 1 & 0 \\
            w & w^2 & w^6 & w^3 & 0 & 1 & 0
        \end{bmatrix}.
    \end{align*}
Via Magma, the code $C$ is easily verified to be an MDS code with parameters $[7,4,4]$.
It is further determined to be a non-GRS code using Algorithm 2.
\end{Example}

\begin{Remark}
    The maximum code length of non-GRS MDS codes in Theorems \ref{leng+} and \ref{leng+4} is equal to that given in \cite[Lemma 4]{Roth}, when $4 \le k \le \frac{q}{2}-1$.
\end{Remark}

As is well-known, when $p = 2$ and $k = 3$, there exists an extended code of an extended GRS code which is a non-GRS MDS code, denoted by $\NGRS_{q+2,3}$.
This code has a length of $q + 2$ and its generator matrix is given by:
\begin{align}\label{}
    \begin{bmatrix}
        1 & 1 & \cdots &1 & 0 & 0 \\
        \alpha_1 & \alpha_2 & \cdots & \alpha_q & 0 & 1 \\
        \alpha_1^2 & \alpha_2^2 & \cdots & \alpha_q^2 & 1 & 0
    \end{bmatrix},
\end{align}
where $\{\alpha_i\}$ runs over all distinct elements in $\mathbb{F}_q$.
The dual code of $\NGRS_{q+2,3}$ has parameters $[q+2, q-1, 4]$.

It is not difficult to see $\NGRS_{q+2,3}$ is a Roth-Lempel code and cannot be constructed via modified GRS codes. This is because both the matrices
$$
\begin{bmatrix}
    1 & 0 & 1 \\
    \eta & 0 & \eta \\
    \eta^2 & 1 & 0
\end{bmatrix}
\ \text{and}\
\begin{bmatrix}
    1 & 0 & 1 \\
    0 & 0 & 0\\
    0 & 1 & \eta
\end{bmatrix}
$$
must be non-singular for any $\eta \in \mathbb{F}_q$.

\begin{Corollary}\label{TLeng}
    If $p=2$, then when $\frac{q}{2} \le k < q-1$, the $\TGRS_{n,k}(\bm{\alpha}, \bm{v}, \lambda, k-1)$
can yield non-GRS MDS code with code length $k+3$.
\begin{proof}
    We consider the punctured codes of the code $\NGRS_{q+2,3}$.
If we puncture any $s$ positions ($s \le q-4$) among its first $q+1$ positions, the resulting punctured code is also a non-GRS MDS code.
Let $\mathcal{N}$ be an arbitrary $(s-1)$-element subset of $\{1, 2, \dots, q\}$.
Then the punctured code of $\NGRS_{q+2,3}$ on $\mathcal{N}\cup\{q+1\}$ has parameters $[q+2-s, 3, q-s]$.
By Proposition \ref{Propcon2} and Corollary \ref{NGR}, its dual code is a $\TGRS_{n,k}(\bm{\alpha}, \bm{v}, \lambda, k-1)$ code with parameters $[q+2-s, q-1-s, 4]$, which is a non-GRS MDS code.
\end{proof}
\end{Corollary}

We present the maximum length of non-GRS (non-elliptic curve) MDS codes as follows.
\begin{table}[h]
\centering
\begin{tabular}{cccc|cccc}
\hline
$p$ & $k$ & $n$ & method  \\
\hline
2 & 3 & $q + 2$ & Roth-Lempel \\
2 & 4 & $\frac{q+6}{2}$ & modified GRS (Theorem \ref{leng+4}) \\
2 & $5 \leq k \leq \frac{q-4}{2}$ & $\frac{q+4}{2}$ & modified GRS (Theorem \ref{leng+}) \\
2 & $\frac{q-2}{2}$ & $\frac{q+6}{2}$  & modified GRS (Theorem \ref{leng+4}) \\
2 & $\frac{q}{2} \leq k < q - 1$ & $k + 3$ & twisted GRS (Corollary \ref{TLeng}) \\
2 & $k = q - 1$ & $q+2$ & dual of Roth-Lempel \\
odd & 3 & $\frac{q+5}{2}$ & modified GRS (Theorem \ref{OddLength3}) &  \\
odd & $4 \leq k \leq \frac{q-3}{2}$ & $\frac{q+3}{2}$ & modified GRS (Theorem \ref{OddLength}) \\
odd & $\frac{q-1}{2}$ & $\frac{q+5}{2}$& modified GRS (Theorem \ref{OddLength3}) \\
\hline
\end{tabular}
\caption{Code length of non-GRS (non-elliptic curve) MDS codes.}
\end{table}

\section{Recovery and determinant algorithm of GRS codes}
In this section, we aim to solve the following two problems:

\begin{Problem}\label{problem1}
    Given an $[n,k]( 3 \le k \le n-3)$ GRS code $C$ over $\mathbb{F}_q$ $(q \ge n)$ whose generator matrix is
    \begin{align}\label{echoleon}
        M=
        \begin{pmatrix}
        1 & 0 & \cdots & 0 & b_{1,k + 1} & \cdots & b_{1,n}\\
        0 & 1 & \cdots & 0 & b_{2,k + 1} & \cdots & b_{2,n}\\
        & & \ddots & & \vdots & & \vdots\\
        0 & \cdots & 0 & 1 & b_{k,k + 1} & \cdots & b_{k,n}
        \end{pmatrix},
    \end{align}
    how can we determine the vectors $\bm{\alpha} = (\alpha_1, \ldots, \alpha_n)$ and $\bm{v}=(v_1, \ldots, v_n)$ such that $C = \GRS_{n,k}(\bm{\alpha}, \bm{v})$?
\end{Problem}

\begin{Problem}\label{problem2}
    Given the generator matrix of a code, how to efficiently determine whether it is an GRS code?
\end{Problem}

\subsection{Transform an extended GRS code to a GRS code}
Here, we present an algorithm for converting an extended GRS code into a GRS code, which mainly leverages Proposition \ref{trans}.  
For this algorithm, we consider a general extended GRS code $\EGRS_{n,k}(\bm{\alpha},\bm{v})$ over $\mathbb{F}_q$,
where one of its evaluation points satisfies $\alpha_t = \infty$, and the code length satisfies $n \le q$.  
The specific steps of the algorithm are as follows.

\noindent\textbf{Step 1}: If there exists $\alpha_j = 0$ for $1 \le j \le n$ with $j\neq t$,
find an element $\lambda \in  \mathbb{F}_q\setminus \{\alpha_i\}$,
    then $$\EGRS_{n,k}(\bm{\alpha}, \bm{v}) = \EGRS_{n,k}((\alpha_1-\lambda,..., \alpha_{t-1}-\lambda,\alpha_t,\alpha_{t+1}-\lambda,...,\alpha_n-\lambda), (v_1,...,v_n)).$$
Note that $\alpha_i-\lambda \neq 0$; we replace $\alpha_i$ by $\alpha_i-\lambda$, for $i = 1,\ldots,t - 1,t + 1,\ldots,n$.

\noindent\textbf{Step 2}: If $\alpha_j \neq 0$ for all $j=1,...,n$,
    then $$\EGRS_{n,k}(\bm{\alpha}, \bm{v}) = \GRS_{n,k}((\alpha_1^{-1},...,\alpha_n^{-1}), (v_1\alpha_1^{k-1},...,v_{t-1}\alpha_{t-1}^{k-1},v_t,v_{t+1}\alpha_{t+1}^{k-1},...,v_{n}\alpha_{n}^{k-1})).$$

We refer to the above two steps as the TransToGRS algorithm,
which has two forms: $\bm{\alpha}, \bm{v} = TransToGRS(\bm{\alpha}, \bm{v})$ and $\bm{\alpha} = TransToGRS(\bm{\alpha})$.
The core function of this algorithm is to convert the vector $\bm{\alpha}$ containing $\infty$ into a new vector $\bm{\alpha}$ without $\infty$.  

\begin{Remark}\label{moveinf}
    For an extended GRS code of length $n = q+1$, its infinite evaluation point $\infty$ can also be moved to the $n$-th coordinate position using a method similar to the above steps,
and the specific process is not elaborated on here.
\end{Remark}

\subsection{Recovery algorithm of GRS codes}
Problem \ref{problem1} stems from the public-key cryptosystem of the Niederreiter scheme \cite{Niederreiter}.
Sidelnikov and Shestakov were the first to successfully launch an attack on the Niederreiter scheme and proposed the Sidelnikov-Shestakov attack method \cite{Sidelnikov}.
This attack method is also elaborated in \cite{wieschebrink2010cryptanalysis}.
In this section, we aim to present a more efficient algorithm to recover the structure of the GRS code from an echelon form matrix as (\ref{echoleon}).

It is an obvious result that the punctured code or shortened code of a GRS code is still a GRS code.
We consider the submatrix $M_2$ of $M$, where
\begin{align}\label{echelonFo}
    M_2=
    \begin{pmatrix}
    1 & 0 & 0 & b_{1,k + 1} & \cdots & b_{1,n}\\
    0 & 1 & 0 & b_{2,k + 1} & \cdots & b_{2,n}\\
    0 & 0 & 1 & b_{3,k + 1} & \cdots & b_{3,n}
    \end{pmatrix}.
\end{align}
So, by Proposition \ref{trans} and Remark \ref{remarkeq}, $M_2$ can be considered as a generator matrix of the extended GRS code
$$\EGRS((\alpha_1,\alpha_2,\alpha_3,\alpha_{k+1},...,\alpha_n),(v_1',v_2',v_3',v_{k+1}',...,v_{n}')).$$
We assume that $\alpha_1 = 0$, $\alpha_2 = 1$, $\alpha_3=\infty$ and $v_1' = 1$.

By row transforming matrix $M_2$ into the form of the generator matrix of an extended GRS code, we obtain
\begin{align*}
    M_2&\rightarrow
    \begin{pmatrix}
    1 & 0 & 0 & b_{1,k + 1} & \cdots & b_{1,n} \\
    0 & v_2' & 0 & v_2'b_{2,k + 1} & \cdots & v_2'b_{2,n} \\
    0 & 0 & v_3' & v_3'b_{3,k + 1} & \cdots & v_3'b_{3,n}
    \end{pmatrix} \\
    &\rightarrow
    \begin{pmatrix}
        1 & v_2' & 0 & b_{1,k + 1}+v_2'b_{2,k + 1} & \cdots & b_{1,n}+v_2'b_{2,n} \\
        0 & v_2' & 0 & v_2'b_{2,k + 1} & \cdots & v_2'b_{2,n} \\
        0 & v_2' & v_3' & v_3'b_{3,k + 1} + v_2'b_{2,k + 1} & \cdots & v_3'b_{3,n} + v_2'b_{2,n}
    \end{pmatrix}.
\end{align*}
Thus
\begin{align*}
    \begin{cases}
        \frac{v_2'b_{2,k + 1}}{b_{1,k + 1}+v_2'b_{2,k + 1}}=\frac{v_3'b_{3,k + 1} + v_2'b_{2,k + 1}}{v_2'b_{2,k + 1}}, \\
        \frac{v_2'b_{2,k + 2}}{b_{1,k + 2}+v_2'b_{2,k + 2}}=\frac{v_3'b_{3,k + 2} + v_2'b_{2,k + 2}}{v_2'b_{2,k + 2}}.
    \end{cases}
\end{align*}
Solving the equation, we have
\begin{align}\label{alpha1}
    \begin{cases}
        v_2'&=\frac{b_{1,k + 2}b_{2,k + 2}b_{1,k + 1}b_{3,k + 1}-b_{1,k + 1}b_{2,k + 1}b_{1,k + 2}b_{3,k + 2}}{b_{1,k + 1}b_{2,k + 1}b_{2,k + 2}b_{3,k + 2}-b_{1,k + 2}b_{2,k + 2}b_{2,k + 1}b_{3,k + 1}}, \\
        \alpha_j&=\frac{v_2'b_{2,j}}{b_{1,j}+v_2'b_{2,j}} \ \text{for} \ j=k+1,...,n.
    \end{cases}
\end{align}

\noindent\textbf{Case 1}: If $k=3$, we have
\begin{align*}
v_3' =-\frac{b_{1,k + 1}v_2'b_{2,k + 1}}{b_{1,k + 1}b_{3,k + 1}+b_{3,k + 1}v_2'b_{2,k + 1}}, \ \text{and} \
v_i' = b_{1,i} + v_2'b_{2,i},\  \text{for} \ i=4,...,n.
\end{align*}
At this moment, we have $\bm{\alpha},\bm{v}=TransToGRS((\alpha_1,...,\alpha_n), (v_1',...,v_n'))$ and $C = \GRS(\bm{\alpha}, \bm{v})$.

\noindent\textbf{Case 2}: If $k>3$.

Consider the $i$-th row ${\bm b}_i$ of $M$ and the associated polynomial $f_{{\bm b}_i}$.
Since the entries $b_{i,1},\ldots,b_{i,i - 1}$ and $b_{i,i+1},\ldots,b_{i,k}$ of ${\bm b}_i$ are equal to zero and $f_{{\bm b}_i}$ has degree at most $k - 1$,
the polynomial must have the form
\begin{align} \label{asspoly}
    f_{{\bm b}_i}(y)=
    \begin{cases}
        c_{b_i}\cdot\prod_{j = 1,j\neq i}^{k}(y - \alpha_j) & \text{if} \ i=3, \\
        c_{b_i}\cdot\prod_{j = 1,j\neq i,j\neq3}^{k}(y - \alpha_j) & \text{if} \ i\neq 3,
    \end{cases}
\end{align}
with $c_{b_i}\in\mathbb{F}_q^*$.
Now pick two arbitrary rows of $M$, for example $\bm{b}_1$ and $\bm{b}_i$,
and divide the entries of the first row by the corresponding entries in the second row as long as these are different from zero.
Using Equation (\ref{asspoly}) we get
\[
    \frac{b_{1,j}}{b_{i,j}}=\frac{v_j\cdot f_{{\bm b}_1}(\alpha_j)}{v_j\cdot f_{{\bm b}_i}(\alpha_j)}=\frac{c_{b_1}(\alpha_j - \alpha_i)}{c_{b_i}(\alpha_j - \alpha_1)}, \ \text{for} \ i=2,4,...,k,j=k+1,...,n.
\]
$$
\begin{cases}
    \frac{b_{1,k+1}}{b_{i,k+1}}(\alpha_{k+1})=\frac{c_{b_1}}{c_{b_i}}(\alpha_{k+1} - \alpha_i), \\
    \frac{b_{1,k+2}}{b_{i,k+2}}(\alpha_{k+2})=\frac{c_{b_1}}{c_{b_i}}(\alpha_{k+2} - \alpha_i).
\end{cases}
$$
Solving the equation, we have
\begin{align}\label{alpha2}
    \alpha_i=\frac{\left(\frac{b_{1,k + 2}}{b_{i,k + 2}}-\frac{b_{1,k + 1}}{b_{i,k + 1}}\right)\alpha_{k + 1}\alpha_{k + 2}}{\frac{b_{1,k + 2}}{b_{i,k + 2}}\alpha_{k + 2}-\frac{b_{1,k + 1}}{b_{i,k + 1}}\alpha_{k + 1}}, i=4,...,k.
\end{align}

Using $\bm{\alpha}=TransToGRS((\alpha_1,\alpha_2,...,\alpha_n))$, we have determined the vector $\bm{\alpha}$ without $\infty$.
Next, we find an alternative $\mathbf{v}$; we may assume that $v_{k+1} = 1$. Let's consider the matrix
\begin{align*}
    \begin{pmatrix}
        1 & 1 &  \cdots & 1 \\
        \alpha_1 & \alpha_2 & \cdots & \alpha_{k+1} \\
        \vdots & \vdots & \ddots & \vdots \\
        \alpha_1^{k-1} & \alpha_2^{k-1} & \cdots &\alpha_{k+1}^{k-1}
    \end{pmatrix}
    \rightarrow
    A=\begin{pmatrix}
        1 & 0 & \cdots & 0 & \frac{g_1(\alpha_{k+1})}{g_1(\alpha_1)} \\
        0 & 1 & \cdots & 0 & \frac{g_2(\alpha_{k+1})}{g_2(\alpha_2)} \\
        & & \ddots &  \vdots  & \vdots \\
        0 & 0 & \cdots & 1 & \frac{g_k(\alpha_{k+1})}{g_k(\alpha_k)}
    \end{pmatrix},
\end{align*}
where $g_i(x)=\prod_{j=1,j\ne i}^{k}(x-\alpha_j)$ for $i = 1,...,k$.
We have
\begin{align}\label{vvv1}
    v_i= \frac{g_i(\alpha_{k+1})}{g_i(\alpha_1)b_{i,k+1}}, \ \text{for} \ i=1,2,...,k,
\end{align}
since $A[i,k+1]\frac{v_{k+1}}{v_i}=b_{i,k+1}$, where $A[i,k+1]$ denotes the entry of matrix $A$ in the $i$-th row and $(k+1)$-th column.

For the $1$st row ${\bm b}_1$ of $M$, we associate it with a new polynomial $h_{{\bm b}_1}$, defined as 
\[
    h_{{\bm b}_1}(y)=d_{b_1}\prod_{j = 2}^{k}(y - \alpha_j), \ \text{where} \ d_{b_1}=\left(v_1\prod_{j = 2}^{k}(\alpha_1 - \alpha_j)\right)^{-1}
\]
since $v_1h_{{\bf b}_1}(\alpha_1)=1$. Then,
\begin{align}\label{vvv2}
    v_i=\frac{b_{1,i}}{h_{{\bm b}_1}(\alpha_i)} \ \text{for} \ i=k+1,...,n, \ \text{since} \ v_ih_{{\bm b}_1}(\alpha_i)=b_{1,i}.
\end{align}
Thus, we have obtained a pair of vectors $\bm{\alpha}=(\alpha_1,\ldots,\alpha_n)$ and $\bm{v}=(v_1,\ldots,v_n)$ such that $C = \GRS(\bm{\alpha}, \bm{v})$.

\begin{Proposition}
    In summary, we obtain the vectors $\bm{\alpha}=(\alpha_1,\alpha_2,\dots,\alpha_n)$ and $\bm{v}=(v_1,v_2,\dots,v_m)$,
such that the GRS code $\GRS(\bm{\alpha}, \bm{v})$ has a generator matrix $M$ (see (\ref{echoleon})).
\end{Proposition}

The corresponding algorithm is presented as follows.
We then estimate the complexity of the above algorithm and compare it with the Sidelnikov-Shestakov attack method.

\begin{algorithm}
    \SetAlgoLined
    \SetKwInOut{Input}{input}\SetKwInOut{Output}{output}
    \Input{$\bm{M}$.}
    \Output{$\bm{\alpha},\bm{v}$.}
    $\alpha_1=0,\alpha_2=1,\alpha_3=\infty$\;
    $v_2'=\frac{b_{1,k + 2}b_{2,k + 2}b_{1,k + 1}b_{3,k + 1}-b_{1,k + 1}b_{2,k + 1}b_{1,k + 2}b_{3,k + 2}}{b_{1,k + 1}b_{2,k + 1}b_{2,k + 2}b_{3,k + 2}-b_{1,k + 2}b_{2,k + 2}b_{2,k + 1}b_{3,k + 1}}$\;
    $\alpha_j=\frac{v_2'b_{2,j}}{b_{1,j}+v_2'b_{2,j}} \ \text{for} \ j=k+1,...,n$\;
    \If{$k == 3$}{
        $v_3' =-\frac{b_{1,k + 1}v_2'b_{2,k + 1}}{b_{1,k + 1}b_{3,k + 1}+b_{3,k + 1}v_2'b_{2,k + 1}}$\;
        $v_i' = b_{1,i} + v_2'b_{2,i},\  \text{for} \ i=4,...,n.$\;
        \Return{$\bm{\alpha},\bm{v}=TransToGRS((\alpha_1,...,\alpha_n), (v_1',...,v_n'))$}\;
    }
    $\alpha_i=\frac{\left(\frac{b_{1,k + 2}}{b_{i,k + 2}}-\frac{b_{1,k + 1}}{b_{i,k + 1}}\right)\alpha_{k + 1}\alpha_{k + 2}}{\frac{b_{1,k + 2}}{b_{i,k + 2}}\alpha_{k + 2}-\frac{b_{1,k + 1}}{b_{i,k + 1}}\alpha_{k + 1}}$, $i=4,...,k$\;
    $\bm{\alpha} = TransToGRS((\alpha_1,...,\alpha_n))$\;
    $
    v_i=-\frac{g_i(\alpha_{k+1})}{g_i(\alpha_i)b_{i,k+1}}, \text{where} g_i(x)= \prod_{j=1,j\ne i}^{k}(x-\alpha_j), \ \text{for} \ i=1,2,...,k 
    $\;
    $d_{b_1}=1/(v_1\prod_{j = 2}^{k}(\alpha_1 - \alpha_j))$,$h_{\bm{b}_1}(y)=d_{b_1}\prod_{j = 2}^{k}(y - \alpha_j)$\;
    $v_i=b_{1,i}/h_{\bm{b}_1}(\alpha_i) \ \text{for}\ i=k+1,...,n.$\;
    \Return{
        $\bm{\alpha}=(\alpha_1,...,\alpha_n),\bm{v}=(v_{1}, v_{2},\ldots,v_{n})$\;
    }
    \caption{Recovery algorithm of an extended GRS code.}
    \label{alg1}
\end{algorithm}

\begin{Remark}
    It is easy to conclude that the computational complexity for calculating the vector $\bm{\alpha}$ in Equations (\ref{alpha1}) and (\ref{alpha2}) is $O(n)$.
The computational complexity for calculating the vector $\bm{v}$ in Equations (\ref{vvv1}) and (\ref{vvv2}) is $O(kn)$.
\end{Remark}

The Sidelnikov-Shestakov attack method described in \cite{wieschebrink2010cryptanalysis} requires two key steps:
firstly, guessing the value of $c_{b_1}/c_{b_2}$ in Equation (\ref{asspoly}),
and then performing operations with a complexity of $O(k^2n + k^3)$ to complete the entire algorithm process.
Therefore, to accurately recover the structure of a GRS code using the Sidelnikov-Shestakov attack method,
it is necessary to guess the value of $c_{b_1}/c_{b_2}$ by traversing all elements in $\mathbb{F}_q^*$,
followed by executing an algorithm with a complexity of $O(k^2n + k^3)$ for recovery.
Subsequently, it is required to verify whether the recovered code is consistent with the original code:
if consistent, the recovery is successful; otherwise, the traversal and guessing process must continue.
The comprehensive computational complexity of this process is much higher than that of Algorithm \ref{alg1}.

\subsection{Determinant algorithm of GRS codes}

Let $G$ be a generator matrix of an $[n,k]$ ($3\le k \le n-3$) linear code $C$.
For Problem \ref{problem2}, since the code $C$ to be determined is not necessarily a GRS code, directly running Algorithm 1 may lead to errors.
Therefore, the following improvements to Algorithm \ref{alg1} are required:
\begin{enumerate}
    \item \label{enum:first} Transform the generator matrix $G$ into an echelon form (see Matrix (\ref{echelonFo}));
    if the transformation fails, then the code $C$ is determined to be a non-GRS code.
    \item \label{enum:second} For all terms involving denominators in Algorithm \ref{alg1},
    it is necessary to check whether they are nonzero;
    if there exists a term with a zero denominator,
    then the code $C$ is determined to be a non-GRS code.
    \item\label{enum:third}  For all $\alpha_i$ generated by Algorithm \ref{alg1},
    it is necessary to verify whether they are pairwise distinct,
    and verify whether all $v_i$ are nonzero;
    if either condition is not satisfied, then the code $C$ is determined to be a non-GRS code.
    \item If Algorithm \ref{alg1} can run normally and generate a GRS code,
    it is necessary to further determine whether the generated GRS code is identical to the original code $C$:
    if they are different, then the code $C$ is a non-GRS code; if they are identical,
    then the code $C$ is a GRS code.
\end{enumerate}

\begin{algorithm}
    \SetAlgoLined
    \SetKwInOut{Input}{input}\SetKwInOut{Output}{output}
    \Input{$\bm{G}$.}
    \Output{$\bm{\alpha},\bm{v},d$.}
    $M, r = Ech(G)$\;
    \tcp{Ech: if success, $M$ is echelon form from $G$ and $r$ is $\True$, if fail $r$ is $\False$.}
    \If{$r == \False$}{
        \Return{$\bm{\alpha}=\bm{0}, \bm{v}=\bm{0}, \False$}\;
    }
    $\bm{\alpha}, \bm{v}, r = ImprovedAlgorithm1(M)$\;
    \tcp{ImprovedAlgorithm1: Obtained by applying the aforementioned improvements \ref{enum:first}, \ref{enum:second}, and \ref{enum:third} to Algorithm \ref{alg1}.}
    \If{$r == \False$}{
        \Return{$\bm{\alpha}=\bm{0}, \bm{v}=\bm{0}, \False$}\;
    }
    $C_1 = EGRSCode(\bm{\alpha}, \bm{v})$,$G_1 = GeneratorMatrix(C_1)$,$M_1, r = Ech(G_1)$\;
    \If{$M_1 == M$}{
        \Return{$\bm{\alpha}, \bm{v}, \True$}\;
    }
    \Return{$\bm{\alpha}=\bm{0}, \bm{v}=\bm{0}, \False$}\;
    \caption{Determinant algorithm of an extended GRS code.}
    \label{alg2}
\end{algorithm}

\begin{Remark}
    For Problem \ref{problem2},
     if the relevant determinant algorithm is implemented using Lemma \ref{cauchy2}, its computational complexity is approximately $O((n-k)^3k^3)$.
    Thus, our Algorithm \ref{alg2} is more efficient.
\end{Remark}

\begin{Remark}
    Based on Remark \ref{moveinf}, Algorithms \ref{alg1} and \ref{alg2} can be easily modified to be applicable to extended GRS codes of length $q+1$.
\end{Remark}

One final interesting problem is as follows: It is well-known that the punctured code and shortened code of an MDS code (or a GRS code) at any position remain an MDS code (or a GRS code), respectively.
A known result states that if the punctured code and shortened code of a code at a certain position are both MDS codes,
then the original code must be an MDS code. Does the GRS code have a similar property? 
That is, we need to investigate the following problem:
\begin{Problem}
If the punctured code and shortened code of a code at a certain position are both GRS codes, must the original code be a GRS code?
\end{Problem}
For this problem, we provide the following counterexample:

\begin{Example}
Let code $C$ be an MDS code over the finite field $\mathbb{F}_{11}$ with parameters $[7,4,4]$,
whose generator matrix is given by
\[            
    G= \begin{pmatrix}
    1 & 0 & 0 & 0 & 1 & 9 & 4 \\
    0 & 1 & 0 & 0 & 4 & 6 & 2 \\
    0 & 0 & 1 & 0 & 5 & 5 & 6 \\
    0 & 0 & 0 & 1 & 6 & 4 & 6
    \end{pmatrix}.
\]
It can be verified via Algorithm \ref{alg2} that the punctured code and shortened code of $C$ at the $7$-th coordinate position are both GRS codes, while $C$ itself is a non-GRS code.
\end{Example}

\vskip 2mm
\noindent\textbf{Acknowledgement.}

This work was supported by
The National Natural Science Foundation of China (Grant Nos. 12271199, 12441102, 12171191)
and the Fundamental Research Funds for the Central Universities grant no. CCNU25JCPT031.

\end{document}